\documentclass[12pt]{article}
\usepackage{eurosym}
\usepackage{amsfonts}
\usepackage{amssymb,amsmath}
\usepackage{graphicx}
\usepackage{young}
\usepackage{youngtab}
\usepackage{color}
\usepackage{subcaption}

\newcommand{\be}{\begin{equation}}
\newcommand{\ee}{\end{equation}}
\newcommand{\bea}{\begin{eqnarray}}
\newcommand{\eea}{\end{eqnarray}}
\newcommand{\beas}{\begin{eqnarray*}}
\newcommand{\eeas}{\end{eqnarray*}}
\newcommand{\ba}{\begin{array}}
\newcommand{\ea}{\end{array}}

\newcommand{\nbox}{{\,\lower0.9pt\vbox{\hrule \hbox{\vrule height 0.2 cm \hskip 0.19 cm \vrule height 0.2 cm}\hrule}\,}}

\def\href#1#2{#2}
\input{tcilatex}

\begin{document}

\begin{titlepage}
\hfill
\vbox{
    \halign{#\hfil         \cr
           } 
      }  

\hbox to \hsize{{}\hss \vtop{ \hbox{}

}}

\begin{flushright}
WITS-MITP-007
\end{flushright}

\vspace*{20mm}
\begin{center}

{\large \textbf{Graph duality as an instrument of Gauge-String correspondence}}

\vspace{ 8mm}

{\normalsize {Pablo D\'iaz${}^{1}$, Hai Lin${}^{2,3,4}$, and Alvaro V\'eliz-Osorio${}^{5}$}  }

{\normalsize \vspace{10mm} }

{\small \emph{${}^1$\textit{Department of Physics and Astronomy, University of Lethbridge, \\
Lethbridge,
Alberta, T1K 3M4, Canada
}} }

{\normalsize \vspace{0.2cm} }

{\small \emph{${}^2$\textit{Department of
Mathematics, Harvard University, Cambridge, MA 02138, USA
}} }

{\normalsize \vspace{0.2cm} }

{\small \emph{${}^3$\textit{Center of Mathematical Sciences and Applications, Harvard University, \\Cambridge, MA 02138, USA
}} }

{\normalsize \vspace{0.2cm} }

{\small \emph{$^4$\textit{Department of
Physics, Harvard University, Cambridge, MA 02138, USA
\\
}} }

{\normalsize \vspace{0.2cm} }

{\small \emph{$^5$\textit{Mandelstam Institute for Theoretical Physics, University of the Witwatersrand,\\ WITS 2050, Johannesburg, South Africa
}} }

{\normalsize \vspace{0.4cm} }

\end{center}

\begin{abstract}

We explore an identity between two branching graphs and propose a physical meaning in the context of the gauge-gravity correspondence. From the mathematical point of view, the identity equates probabilities associated with $\mathbb{GT}$, the branching graph of the unitary groups, with probabilities associated with $\mathbb{Y}$, the branching graph of the symmetric groups. In order to furnish the identity with physical meaning, we exactly reproduce these probabilities as the square of three point functions involving certain hook-shaped backgrounds. We study these backgrounds in the context of LLM geometries and discover that they are domain walls interpolating two AdS spaces with different radii. We also find that, in certain cases, the probabilities match the eigenvalues of some observables, the embedding chain charges. We finally discuss a holographic interpretation of the mathematical identity through our results.

\end{abstract}

\end{titlepage}

\tableofcontents
\newpage

\section{Introduction}

\label{sec: introduction}

The proposal of the AdS/CFT correspondence \cite%
{Maldacena:1997re,Gubser:1998bc,Witten:1998qj} between type II superstring
theory living in $AdS_{5}\times S^{5}$ space and $\mathcal{N}=4$ gauge
theory in four dimensions offers new ways to tackle problems in physics.
Type II superstring theory is a theory of closed strings. It is a UV
completion of gravity. In this correspondence, all the states and dynamics of one theory can be translated into the
other if one has the dictionary. One of the most appealing
applications of the correspondence is to understand features of quantum
gravity by means of studying the gauge theory side.

In fact, since the correspondence was proposed there has been much progress in
identifying features of quantum gravity from the field theory side.
Gravitons as well as branes and spacetime geometries can be seen as emergent
phenomena from various operators in the gauge theory side. It can be shown
that single-particle Kaluza-Klein gravitons in spacetime correspond to gauge
invariant single trace operators in the dual gauge theory. Using spin chains
the spectrum of the string rotating with a large angular momentum agrees
with the states of the field theory produced by composite operators with
scaling dimension of $O(\sqrt{N})$ \cite{Berenstein:2002jq,Minahan:2002ve,Beisert:2003tq,Beisert:2003yb}.
The action of the dilatation operator has also been computed \cite{Berenstein:2006qk,de Mello
Koch:2007uu,Correa:2006yu,de Mello Koch:2007uv,Bekker:2007ea} for the open spin chains and the open
strings. Extended objects like Giant Gravitons \cite%
{McGreevy:2000cw,Grisaru:2000zn,Hashimoto:2000zp,Corley:2001zk} have been
identified with composite operators with scaling dimension of $O(N)$ in the
field theory side. The dynamics of Giant Gravitons from the field theory
side has also been computed by using Young diagrams and Schur technology \cite%
{Koch:2011hb,Carlson:2011hy,Koch:2012sf,deMelloKoch:2011vn,deMelloKoch:2012ck,deMelloKoch:2011ci,Balasubramanian:2001nh}.

Despite the progress and evidence of its validity from non-trivial tests,
the problem of proving the correspondence is still tough. The main reason is that
the weak/strong coupling nature of the duality makes it difficult to
study perturbative regimes of both theories at the same time. The weak
coupling regime of string theory, where at low energies one recovers
low-energy supergravity, corresponds to strong coupling regime of the field
theory where calculations are difficult, and vice versa. However, good understanding of how the duality works can be gained by
studying the underlying mathematical structures and the connections they imply.
Understanding those connections often leads to new insights and results. For
instance, an important mathematical equivalence, Schur-Weyl duality, has been
proved to be behind the map between gauge theory states and stringy
spacetime states \cite%
{Corley:2001zk,Ramgoolam:2008yr,Brown:2007xh,Kimura:2007wy,Kimura:2008ac,Bhattacharyya:2008rb}.

In this paper we go a step further in this line of thinking. Quite recently,
Borodin and Olshanski (BO) \cite{BO} have found interesting identities between the
Gelfand-Tsetlin graph $\mathbb{GT}$ (which corresponds to unitary groups) and the Young
graph $\mathbb{Y}$ (that is associated to the symmetric groups). As said above, it is believed that the AdS/CFT
correspondence has deep roots in the connections between unitary and symmetric groups. This suggests that the identities found by Borodin and Olshanski could have a gauge-gravity interpretation. The task of this paper is to investigate it.
Specifically, we study the BO identity
\begin{equation}
\lim_{\frac{N}{M}\rightarrow \frac{r}{r^{\prime }}}{}^{\mathbb{GT}}\Lambda
_{N}^{M}([\mu ,M],[\nu ,N])={}^{\mathbb{YB}}\Lambda _{r}^{r^{\prime }}(\mu
,\nu ),  \label{sim_0}
\end{equation}%
where $M,N$ with $M>N$ are large. As we will explain, this identity relates probabilities naturally associated to the branching graph of unitary groups (LHS) with those associated to the Young Bouquet (RHS), which is a composite system that contains the branching graph of symmetric groups.

The main results of this paper are the following. The RHS of (\ref{sim_0}) is exactly reproduced at the end of section \ref{sec: multi-graviton transitions}, coming from three point functions that involve excitations over certain hook-shaped backgrounds. For large $N$ and $M$, as the BO identity requires, those hook-shaped backgrounds can be identified with two-ring bubbling geometries \cite{Lin:2004nb} as shown in  section \ref{sec: multi-ring and geometries}. Moreover, in the same section these geometries are shown to interpolate between two AdS spaces with different radii. On the other hand, the LHS of (\ref{sim_0}) coincides, in the case that $\mu=\nu$, with the eigenvalues of the embedding charges constructed in \cite{Diaz:2013gja,Diaz:2014ixa}, a fact which is explained in section \ref{sec: conserved charges}.

These results suggest a holographic interpretation. We go through it in subsection \ref{holointer}. We have related the RHS of (\ref{sim_0}) to the probability of the processes described by the three point functions. We interpret the dual of these processes in the spacetime as gravity processes where an excitation $\mu$ propagates from a region of the two-ring geometry that approaches an AdS region, to an excitation $\nu$ on the other AdS region. It would be interesting to confirm this picture by a direct spacetime computation, but this is out of the scope of this article. Moreover, the fact that the geometry is a domain wall that interpolates between two different AdS is suggestive of an interpretation via an RG flow in the field theory side \cite{Girardello:1998pd,Bianchi:2001de}. The interpretations along these lines would also give a meaningful picture of the embedding charges mentioned above.

The paper is organized as follows. Section \ref{sec: GT and Y graphs} is a
compendium of the mathematical tools which are necessary to understand the
identity (\ref{sim_0}) from the mathematical point of view. Thus, we
explain in section \ref{sec: GT and Y graphs} what the Young graph, the
Gelfand-Tsetlin graph, and the Young Bouquet are. We also see how probability
distributions are naturally assigned to those systems. We refer the interested
reader to the paper by Borodin and Olshanski \cite{BO} for a more comprehensive
treatment of the topics. In section \ref{sec: multi-graviton transitions} we
compute the three-point functions with the hook-shaped background states.
The transition probabilities on hook-shaped background states are given by
a sum of the squares of three-point functions. We see that in a natural limit, these
probabilities match the RHS of (\ref{sim_0}). The interpretation of the LHS of (\ref%
{sim_0}) in terms of conserved charges is explained in section \ref{sec:
conserved charges}. We will see the precise connection of the LHS of (\ref%
{sim_0}) with the eigenvalues of the embedding charges and draw some
conclusions upon it. Section \ref{sec: multi-ring and geometries} is devoted
to an analysis of the hook-shaped background states and their relation with
bubbling geometries, and also a discussion on the dual spacetime picture of
the above transition processes. We first study multi-ring geometries in the
phase space plane and find the connection between the different radii of the
rings and the rank of the gauge groups in an embedding chain. Then we focus on
the two-ring geometry (hook-shaped) to demonstrate that the background
interpolates between two different AdS spaces. Then, in section \ref{holointer}
we give a possible holographic explanation of the transition processes. In section
\ref{sec: compatibility condition} it is seen that a ring-splitting process is
related to the compatibility condition of the probabilities in the
Gelfand-Tsetlin graph. Finally, we reserve section \ref{sec: discussion} for
a discussion and an outline of possible future works. In appendix
\ref{sec: four point and multigraviton}, we go beyond the three-point functions
and consider four-point functions in hook-shaped backgrounds, and find nice simple
formulas for a large $N$ limit.

\section{GT graph, Young graph and BO identity}

\label{sec: GT and Y graphs}

In this work we wish to elucidate, in the context of the gauge/string duality, a novel connection between two kinds of leveled graphs. On the one hand, we have the Young graph $\mathbb{Y}$ describing the branching of symmetric groups while on the other we have the Gelfand-Tsetlin graph $\mathbb{GT}$ describing that of unitary groups. Recently, a beautiful relationship between these two graphs has been discovered by Borodin and Olshanski \cite{BO}; in some sense, their result can be viewed as an extension of the celebrated Schur-Weyl duality. As pointed out in \cite{Ramgoolam:2008yr}, the Schur-Weyl duality is a useful tool in the understanding of the gauge/string correspondence, and it is in this spirit that we wish to explore the consequences of the Borodin-Olshanski (BO) identity. In the present section, after introducing $\mathbb{Y}$ and $\mathbb{GT}$, we state the BO identity.

\subsection{The Young graph}

We start by describing the Young graph $\mathbb{Y}$, this is a leveled graph whose vertices correspond to Young diagrams and its  leveling criterion is the number of boxes in each diagram, that is, at level one we have all the Young diagrams with one box, at level two those with two and so on. Clearly, this graph is infinite since it is possible to construct Young diagrams with an arbitrary number of boxes. Vertices in $\mathbb{Y}$ are linked if and only if their corresponding Young diagrams can be obtained from each other by adding or removing a single box, hence links connect only consecutive strata.

If we recall that Young diagrams with $n$ boxes characterize irreducible representations (irreps) of the symmetric group $S_n$, then we can give a group-theoretic interpretation to $\mathbb{Y}$; namely, the Young graph represents how irreps of $S_n$ are subduced by irreps of $S_{n+1}$ for each level $n$. Hereafter, we will reserve the letters $m$ and $n$ to label the levels on this graph, while the letters $\mu$ and $\nu$ will stand for Young diagrams.

From any given vertex $\mu $ in $\mathbb{Y}$
it is possible to follow at least one path downwards all the way to the bottom of the graph. Every such path is a way of decomposing the Young diagram $\mu$ one box at a time. In group theory terminology, each of these paths consists a of list of linked irreps associated with the chain of embeddings:
\begin{equation}
S_{n}\supset S_{n-1}\supset \cdots \supset S_{1}.
\end{equation}%
It can be shown that the number of paths descending from a vertex $\mu$ equals the dimension of the irrep $\mu$, thus each path corresponds to a state of $\mu$. Alternatively, the dimension of $\mu$ can be computed by means of the so-called hook lengths of $\mu$. Let's remind ourselves how this is done. Recall that if $(i,j)$  is a cell in $\mu$, then its \textit{hook} is the set
\begin{equation}
\text{H}_{\mu}(i,j)=\{(a,b)\in \mu| a=i, \, b\geq j\}\cup \{(a,b)\in \mu| b=j, \, a\geq i\},
\end{equation}
and the cell's \textit{hook length} is nothing but $h_{\mu}(i,j)\equiv |\text{H}_{\mu}(i,j)|$. If we define the hook length of the diagram $\mu$ as \footnote{Frequently, the notation $\text{Hooks}_{\mu}$ is used for $H_{\mu}$, and we choose the latter to avoid long expressions in the following sections.}
\begin{equation}\label{hook length}
 H_{\mu}= \prod_{(i,j)\in\mu} h_{\mu}(i,j),
\end{equation}
we can show that
\begin{equation}\label{dim hook}
\text{dim}_{\mu }=\frac{m!}{H_{\mu}}.
\end{equation}%
The above expression will be very useful in the following sections.
Another notion that will be relevant to our discussion is that of the relative dimension $\text{dim}(\mu ,\nu )$ of two vertices $\mu$ and $\nu$, which corresponds to the number of paths descending from $\mu$ to $\nu$.\newline
\newline
\newline
\newline
\newline
\setlength{\unitlength}{2.7cm} {\small
\begin{picture}(3,1)\label{1}
\put(0,1.1){$n=3$}
\put(1.5,1){$\begin{Young}
&\cr
\cr
\end{Young}$}
\put(.7,1){$\begin{Young}
\cr
\cr
\cr
\end{Young}$}
\put(2.5,1){$\begin{Young}
&&\cr
\end{Young}$}
\put(0,.5){$n=2$}
\put(1,.5){$\begin{Young}
\cr
\cr
\end{Young}$}
\put(2,.5){$\begin{Young}
&\cr
\end{Young}$}
\put(.95,.82){\line(-1,1){.14}}
\put(2.3,.68){\line(1,1){.27}}
\put(1.5,.95){\line(-1,-1){.3}}
\put(1.7,1){\line(1,-1){.32}}
\put(0,-.2){$n=1$}
\put(1.5,-.2){$\begin{Young}
\cr
\end{Young}$}
\put(1.55,0){\line(-1,1){.4}}
\put(1.65,0){\line(1,1){.45}}
\end{picture}} \newline
\newline
\newline

Let $\mathbb{G}$ be an arbitrary leveled graph, for any pair of vertices $x,y\in \mathbb{G}$
such that $m>n$ we can define the quantity
\begin{equation}\label{proba}
{}^{\mathbb{G}}\Lambda _{n}^{m}(x ,y )\equiv\left(\frac{\text{\# paths from $y$ to the ground floor}}{\text{\# paths from $x$ to the ground floor}}\right)\times\left(\text{\# paths from $x$ to $y$}\right),
\end{equation}%
which satisfies the property
\begin{equation}\label{nor}
\sum_{y}{}^{\mathbb{G}}\Lambda _{n}^{m}(x ,y )=1,
\end{equation}%
where the sum is over all the vertices $y$ at level $n$. Therefore, it is clear that for any vertex $x\in \mathbb{G}$, the quantity \eqref{proba} furnishes a probability distribution on each level $n<m$  in the graph. Moreover, these distributions satisfy the compatibility condition
\begin{equation}\label{compa}
\sum_{y'}{}^{\mathbb{G}}\Lambda _{n^{\prime
}}^{m}(x ,y ^{\prime }){}^{\mathbb{G}}\Lambda _{n}^{n^{\prime }}(y
^{\prime },y )={}^{\mathbb{G}}\Lambda _{n}^{m}(x,y )
\end{equation}
for the intermediate levels. To lighten the presentation, we express this condition with the shorthand notation
\begin{equation}\label{compa short}
{}^{\mathbb{G}}\Lambda _{n^{\prime }}^{m} \;{}^{\mathbb{G}}\Lambda
_{n}^{n^{\prime }}={}^{\mathbb{G}}\Lambda _{n}^{m}.
\end{equation}
Observe that the above construction is valid for any leveled graph.

Since $\mathbb{Y}$ is a leveled graph, we can associate distributions of type \eqref{proba} to it. In terms of the dimensions, these can be expressed as
\begin{equation}\label{prob young}
{}^{\mathbb{Y}}\Lambda _{n}^{m}(\mu ,\nu )= \frac{\text{dim}_{\nu }%
}{\text{dim}_{\mu }} \text{dim}(\mu ,\nu ).
\end{equation}

Below, we will also be interested in restrictions of the form $S_{n}\times
S_{m-n}\subset S_{m}$, as opposed to $S_{n}\subset S_{m}$ discussed above. The number of times an irrep $(\nu,\nu')$ of $S_{n}\times S_{m-n}$ appears in the restriction of the irrep $\mu$ of $ S_{m}$ is given by the \textit{Littlewood-Richardson coefficients} $g(\mu ;\nu ,\nu
^{\prime })$. These coefficients satisfy the relationship
\begin{equation}
\text{dim}(\mu ,\nu )=\sum_{\nu ^{\prime }\vdash m-n}g(\mu ;\nu ,\nu
^{\prime })\text{dim}_{\nu ^{\prime }}.
\end{equation}

\subsection{The Gelfand-Tsetlin graph}

Now we turn our attention to the study of the Gelfand-Tsetlin graph $\mathbb{GT}$. While the Young graph $\mathbb{Y}$ dealt with irreps of the symmetric group $S_n$,  $\mathbb{GT}$ does so for the unitary groups $U(N)$. The vertices of $\mathbb{GT}$ correspond to irreps of $U(N)$ leveled by the rank $N$. Remember that the irreps of $U(N)$ can also be labeled by Young diagrams. Specifically, level $N$ is conformed by all the Young diagrams with at most $N$ rows; clearly, since there is no bound on the number of columns, there is an infinite number of vertices at each level. Moreover, each level contains all the diagrams present in the levels below. Hence, when speaking of a Young diagram as a vertex in $\mathbb{GT}$ one must always specify the level in question, for example $\left(\mu, N\right)\in\mathbb{GT}$.

We know the leveling criterion and the vertices of $\mathbb{GT}$, we are missing only the links in order to describe the graph completely. To establish whether two vertices are linked in this graph is less straightforward than for $\mathbb{Y}$ and it is necessary to introduce certain preliminary concepts. The \textit{signature} of a vertex $\left(\mu, N\right)\in\mathbb{GT}$ is a $N$-tuple of integers, where the first $k$ numbers ($k\leq N$ is the number of rows of $\mu $) are the lengths of the rows of $\mu $ and the rest are 0's, for
example
\begin{equation}
\bigg(\,\begin{Young} &\cr \cr \end{Young},5\,\bigg)\longleftrightarrow
(2,1,0,0,0).
\end{equation}
We say that the signatures of two vertices in $\mathbb{GT}$, $(r_{1},r_{2},\dots ,r_{N})$ and $(s_{1},s_{2},\dots ,s_{N-1})$ at levels $N$ and $N-1$, respectively, \textit{interlace} if and only if
\begin{equation}
r_{1}\leq s_{1}\leq r_{2}\leq s_{2}\leq \cdots \leq r_{N-1}\leq s_{N-1}\leq
r_{N}.
\end{equation}
Vertices in the Gelfand-Tsetlin graph are linked if their signatures interlace.

Once again, links form paths in this graph and
as you follow the links all the way to the bottom you move through the restriction chain:
\begin{equation}
U(N)\supset U(N-1)\supset \cdots \supset U(1).  \label{Us}
\end{equation}
Similarly to $\mathbb{Y}$, the number of paths from irrep $(\mu ,N)\in\mathbb{GT}$  to the ground floor matches the dimension of the irrep, $\text{Dim}[\mu ,N]$. Also, we define the relative dimension
 $\text{Dim}([\mu ,M],[\nu ,N])$ that corresponds to the number of paths (if any) that join
irrep $[\mu ,M]$ with irrep $[\nu ,N]$ in the graph, with $M>N$.
 Alternatively, the dimensions of irreps of $U(N)$ can be extracted combinatorially from
\begin{equation}
\text{Dim}[\nu ,N]=f_{\nu }(N)\frac{\text{dim}_{\nu }}{n!},\quad |\nu |=n,
\end{equation}%
where
\begin{equation}
f_{\nu }(N)=\prod_{i,j}(N-i+j),
\end{equation}%
is a product over all the cells in $\nu$ and $\text{dim}_{\nu }$ can be found in Eq. \eqref{dim hook}.

Every descending path in $\mathbb{GT}$ can be
represented by a so-called \textit{Gelfand-Tsetlin pattern}, which are a clever way of organizing the signatures of the vertices. We illustrate this with an example.
Consider the vertex
\begin{equation}
\bigg(\,\begin{Young}&\cr
\cr \end{Young},3\,\bigg),
\end{equation}
whose signature is $(2,1,0)$. The valid Gelfand-Tsetlin
patterns are eight in this case (The rows are the signatures of the irreps):
\begin{eqnarray}
&\left(
\begin{array}{ccccc}
2 &  & 1 &  & 0 \\
& 2 &  & 1 &  \\
&  & 2 &  &
\end{array}%
\right) &,\left(
\begin{array}{ccccc}
2 &  & 1 &  & 0 \\
& 2 &  & 1 &  \\
&  & 1 &  &
\end{array}%
\right) ,\left(
\begin{array}{ccccc}
2 &  & 1 &  & 0 \\
& 2 &  & 0 &  \\
&  & 2 &  &
\end{array}%
\right) ,\left(
\begin{array}{ccccc}
2 &  & 1 &  & 0 \\
& 2 &  & 0 &  \\
&  & 1 &  &
\end{array}%
\right) ,  \notag \\
&\left(
\begin{array}{ccccc}
2 &  & 1 &  & 0 \\
& 2 &  & 0 &  \\
&  & 0 &  &
\end{array}%
\right) ,&\left(
\begin{array}{ccccc}
2 &  & 1 &  & 0 \\
& 1 &  & 1 &  \\
&  & 1 &  &
\end{array}%
\right) ,\left(
\begin{array}{ccccc}
2 &  & 1 &  & 0 \\
& 1 &  & 0 &  \\
&  & 1 &  &
\end{array}%
\right) ,\left(
\begin{array}{ccccc}
2 &  & 1 &  & 0 \\
& 1 &  & 0 &  \\
&  & 0 &  &
\end{array}%
\right) .  \notag \\
&&
\end{eqnarray}%
Note that the rule is that in each level down, the numbers must be in
between as the interlace condition dictates. As described above, each GT
pattern is a path in $\mathbb{GT}$:\newline
\newline
\newline

\setlength{\unitlength}{2.7cm} {\small
\begin{picture}(3,1)
\put(0,1.1){$N=3$}
\put(1.5,1){$\begin{Young}
&\cr
\cr
\end{Young}$}
\put(0,.5){$N=2$}
\put(1,.5){$\begin{Young}
\cr
\end{Young}$}
\put(1.5,.5){$\begin{Young}
&\cr
\end{Young}$}
\put(2.2,.5){$\begin{Young}
\cr
\cr
\end{Young}$}
\put(2.7,.5){$\begin{Young}
&\cr
\cr
\end{Young}$}
\put(1.5,.95){\line(-1,-1){.3}}
\put(1.6,.95){\line(0,-1){.25}}
\put(1.7,1){\line(1,-1){.35}}
\put(1.8,1.1){\line(3,-1){.8}}
\put(0,-.2){$N=1$}
\put(1,-.2){$\emptyset$}
\put(1.05,0){\line(0,1){.4}}
\put(1.1,0){\line(1,1){.4}}
\put(1.5,-.2){$\begin{Young}
\cr
\end{Young}$}
\put(1.55,0){\line(-1,1){.4}}
\put(1.57,0){\line(0,1){.4}}
\put(1.65,0){\line(1,1){.5}}
\put(1.67,0){\line(2,1){1}}
\put(2,-.2){$\begin{Young}
&\cr
\end{Young}$}
\put(2.1,0){\line(-2,3){.27}}
\put(2.3,0){\line(1,1){.4}}
\put(3,1.3){$\frac{\text{Dim}({\tiny\yng(1,1)}
,2)}{\text{Dim}({\tiny\yng(2,1)}
,3)}$}
\qbezier(2.9,1.2)(1.2384,1.0)
(1.9,0.7722)
\end{picture}} \newline
\newline
\newline

Since $\mathbb{GT}$ is a leveled graph, probabilities of the form \eqref{proba} can be associated to it. Thus, we have
\begin{equation} \label{prob gt}
{}^{\mathbb{GT}}\Lambda _{N}^{M}(\mu ,\nu )= \frac{\text{Dim}[\nu ,N]}{\text{Dim}[\mu ,M]}\text{Dim}([\mu ,M],[\nu ,N]).
\end{equation}%
It must be clear that ${}^{\mathbb{GT}}\Lambda _{N}^{M}(\mu ,\nu )$  satisfies the normalization condition \eqref{nor} and compatibility condition  \eqref{compa}. We express the latter as
\begin{equation}
{}^{\mathbb{GT}}\Lambda _{N^{\prime }}^{M}{}^{\mathbb{GT}}\Lambda
_{N}^{N^{\prime }}={}^{\mathbb{GT}}\Lambda _{N}^{M}, \label{compa gt}
\end{equation}%
with the shorthand notation of Eq. \eqref{compa short}.

\subsection{The Young Bouquet and the BO identity}

In the previous sections we have described two leveled graphs and their associated probability distributions. These two graphs have some similarities but as a matter of fact they are describing rather different mathematical objects. One might wonder whether there is any quantitative relationship between them. This question was addressed by Borodin and Olshanski \cite{BO} by comparing the probability distributions \eqref{prob young} and \eqref{prob gt}. More precisely, they compared the $\mathbb{GT}$-distribution and a modified version of $\mathbb{Y}$-distribution which we introduce now. A \textit{binomial projective system} is the family of probability distributions
\begin{equation}
{}^{\mathbb{B}}\Lambda _{r}^{r^{\prime }}(m,n)=\Big(1-\frac{r}{r^{\prime }}%
\Big)^{m-n}\Big(\frac{r}{r^{\prime }}\Big)^{n}\frac{m!}{(m-n)!n!},
\label{binomial}
\end{equation}%
where $r,r^{\prime }\in\mathbb{R}^+$  and $n,m$ are non-negative
integers. By combining \eqref{binomial} with \eqref{prob young}, Borodin and Olshanski defined the \textit{Young
Bouquet} whose associated distribution reads
\begin{equation}
{}^{\mathbb{YB}}\Lambda _{r}^{r^{\prime }}(\mu ,\nu )=\Big(1-\frac{r}{%
r^{\prime }}\Big)^{m-n}\Big(\frac{r}{r^{\prime }}\Big)^{n}\frac{m!}{(m-n)!n!}%
\frac{\text{dim}_{\nu }}{\text{dim}_{\mu }} \text{ dim}(\mu ,\nu ),\label{YB dist}
\end{equation}%
where in the above $|\mu |=m$ and $|\nu |=n$, and $m\geq n$. One can check that the
compatibility condition
\begin{equation}
{}^{\mathbb{YB}}\Lambda _{r^{\prime \prime }}^{r^{\prime }}{}^{\mathbb{YB}%
}\Lambda _{r}^{r^{\prime \prime }}={}^{\mathbb{YB}}\Lambda _{r}^{r^{\prime }}
\label{YBcom}
\end{equation}%
holds, where we used the shorthand notation \eqref{compa short}.

It is this object, the Young Bouquet, which is found to have a deep
connection with the Gelfand-Tsetlin graph. The identity found by Borodin and Olshanski is \cite{BO}
\begin{equation}
\lim_{\frac{N}{M}\rightarrow \frac{r}{r^{\prime }}}{}^{\mathbb{GT}}\Lambda
_{N}^{M}([\mu ,M],[\nu ,N])={}^{\mathbb{YB}}\Lambda _{r}^{r^{\prime }}(\mu
,\nu ),  \label{sim}
\end{equation}%
where $N,M\rightarrow \infty $, $M>N$. Formula (\ref{sim}), or its explicit
form {\small
\begin{equation}
\lim_{\frac{N}{M}\rightarrow \frac{r}{r^{\prime }}}\frac{\text{Dim}[\nu ,N]%
}{\text{Dim}[\mu ,M]} \text{Dim}([\mu ,M],[\nu ,N])=\binom{m}{n}\Big(1-\frac{r%
}{r^{\prime }}\Big)^{m-n}\Big(\frac{r}{r^{\prime }}\Big)^{n}\frac{\text{dim}%
_{\nu }}{\text{dim}_{\mu }} \text{ dim}(\mu ,\nu ),  \label{exp}
\end{equation}}%
is a deep mathematical identity which depends only on how the branching
graphs and their boundaries ($N\rightarrow \infty$) are constructed which,
in the end, depends on how irreps of the groups are subduced  \cite{BO,BO2012}. In the following,
we refer to Eq. \eqref{sim} as the BO identity or $\mathbb{YB}/\mathbb{GT}$ duality.

\subsection{Relation to gauge-gravity correspondence}

In the forthcoming sections, we will show that the RHS of the $\mathbb{YB}/\mathbb{GT}$ duality \eqref{sim} can be encoded in a well-defined physical process. The general reasoning goes along the following lines. Young diagrams with at most $N$ rows furnish the entire half-BPS sector of $\mathcal{N}=4$ gauge theory with unitary groups $U(N)$ \cite{Corley:2001zk}. Thus we will argue that the LHS of \eqref{sim} can be understood as a transition probability between states in theories with gauge groups $U(M)$ and $U(N)$ respectively. Hence, applying the AdS/CFT correspondence we can give an interpretation to the RHS of \eqref{sim} in terms of probabilities of transitions in quantum gravity and string theory. As a matter of fact, the aforementioned transitions will correspond to transitions of multigraviton states on certain spacetime backgrounds.

\begin{figure}
        \centering
        \begin{subfigure}[h]{0.5\textwidth}
\includegraphics[scale=0.30,bb=0 0 300 570]{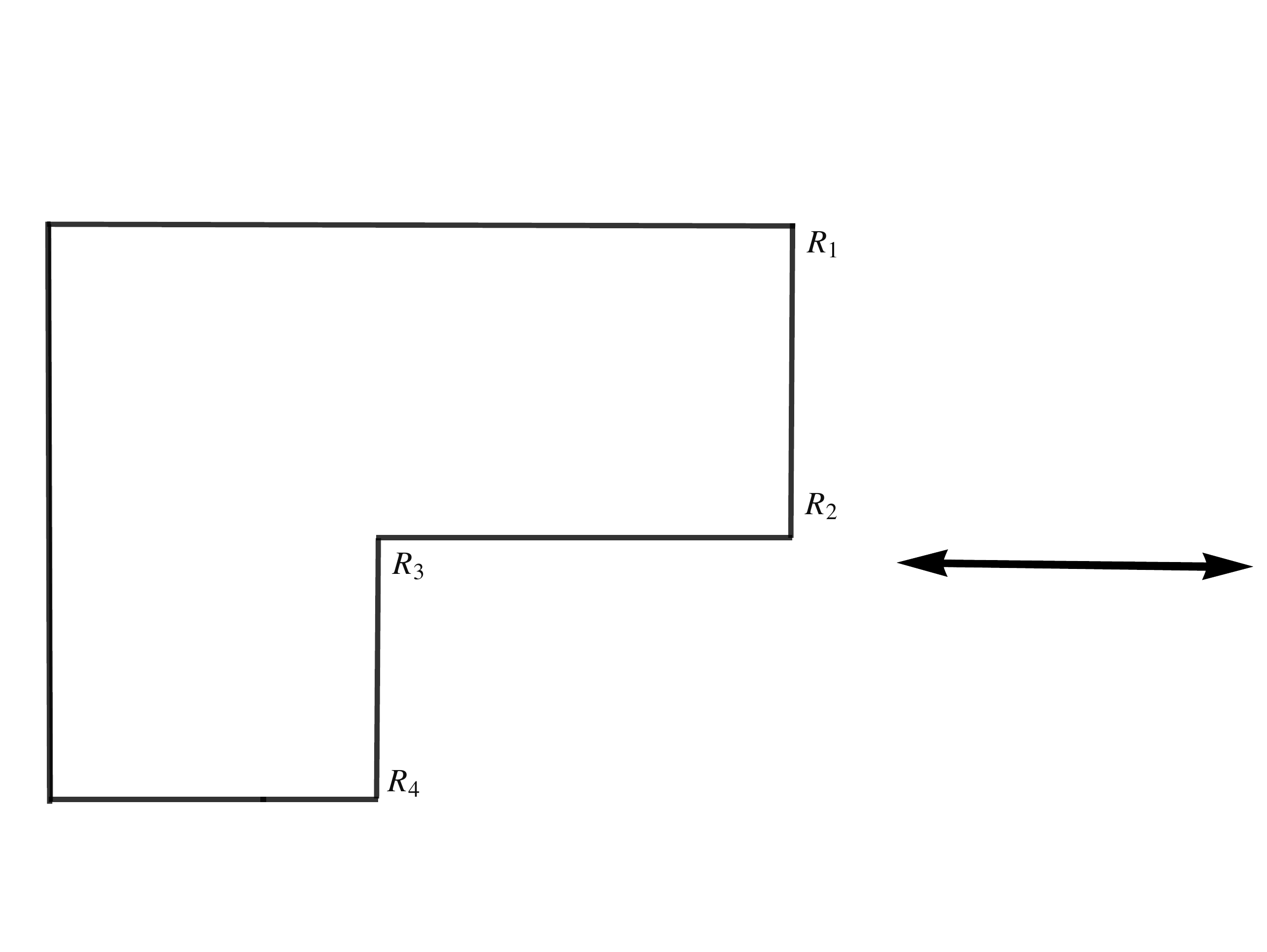}
                      \end{subfigure}%
        ~
    \begin{subfigure}[h]{0.5\textwidth}
\hspace{8mm}\includegraphics[scale=0.32,bb=0 0 300 570]{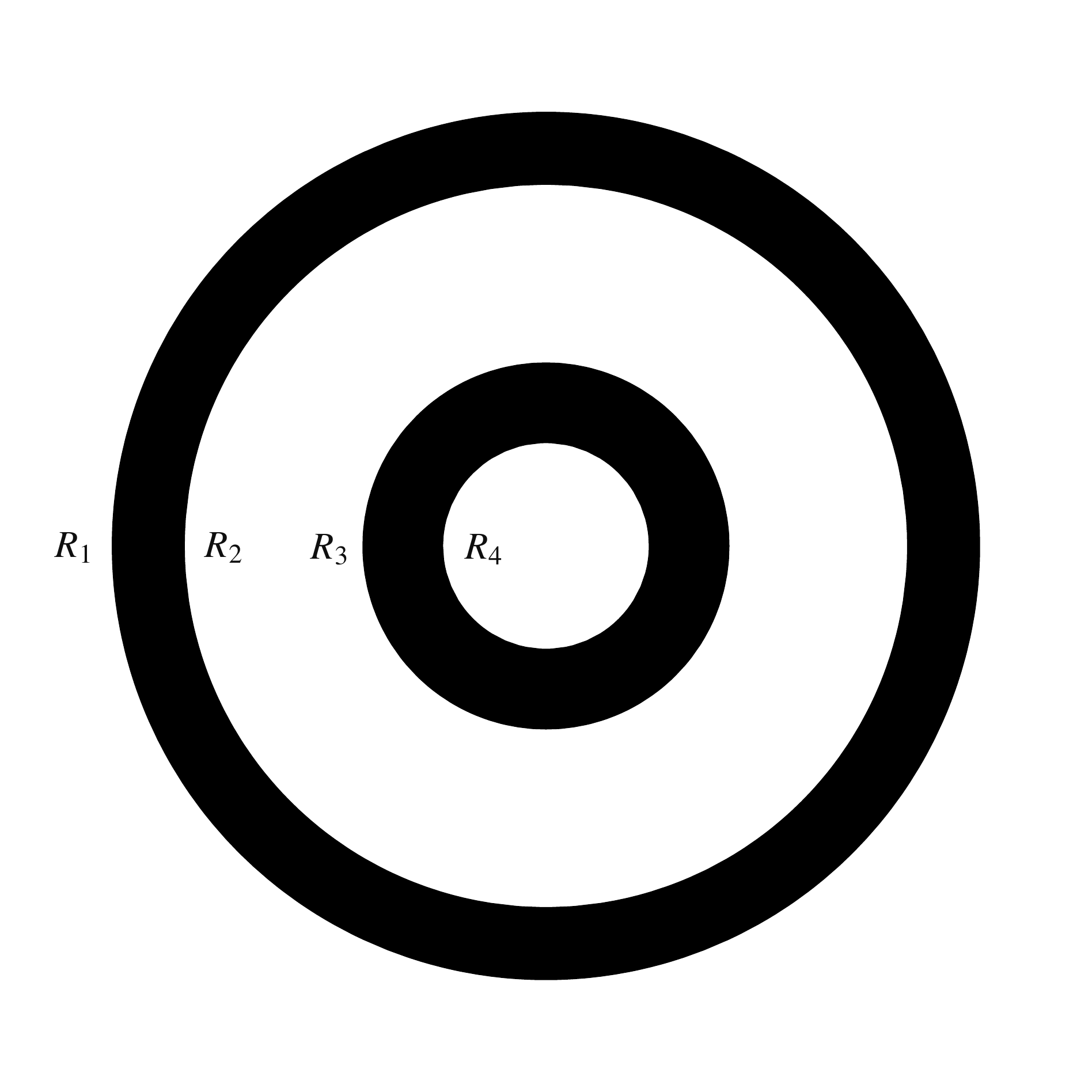}
        \end{subfigure}
        \caption{One-to-one relation between Young diagrams and the bubbling plane.}\label{Young bubbles}
\end{figure}

The backgrounds that are relevant for our discussion can be produced using the LLM prescription \cite{Lin:2004nb}, which allows us to construct the geometries corresponding to half-BPS states explicitly. In principle we could use other kind of backgrounds but for convenience we focus on LLM bubbling geometries.
Let us remind the reader what is the physical meaning behind this construction. First, to a half-BPS state one associates a Young diagram, then from this diagram one constructs a black and white pattern on a plane. For example, in Fig. \ref{Young bubbles} the diagram in the LHS gives rise to two concentric black rings in a sea of white. For a Young diagram with a large number of boxes, this pattern in the so-called \textit{bubbling plane} provides all the information necessary to construct the ten dimensional geometry corresponding to the half-BPS state in the gravity side. In section \ref{sec: multi-ring} we will construct explicitly the background geometries in which the multigravitons must scatter in order to produce the RHS of \eqref{sim}. We will find that they correspond to domain walls that interpolate between two AdS spaces whose radii satisfy
\begin{equation}
\frac{r}{r^{\prime }}=\Big(\frac{R_{AdS}}{R_{AdS}^{\prime }}\Big)^{4},
\end{equation}
where $r$ and $r'$ are precisely those appearing in \eqref{sim}.

Let us make a physical comment about irreps $[\mu ,N]$ of the unitary groups, in this context. It is known that by means of Schur polynomials \cite{Corley:2001zk}, irreps of the unitary group label half-BPS operators of the CFT's\footnote{And hence states, since for CFT's there is a one-to-one correspondence between operators and states.}. It is not clear, though, what the physical role (if any) that the internal states of the irrep play. The internal states of the irreps are identified with the paths in $\mathbb{GT}$. Now, do they have any physical meaning? So far, the paths of the irreps had come into formulas as $\text{Dim}[\mu ,N]$, that is, as its total number. However, in the LHS of (\ref{sim}) we see that the relative dimension, which comes up as we partition the space of paths, is relevant. We claim that the probabilities associated to the transitions between states $\mu$ and $\nu$ in $U(M)$ and $U(N)$ gauge theories respectively, give a novel physical content to the internal states of the irreps of unitary groups.

\section{Three point functions and Young Bouquet distribution}

\label{sec: multi-graviton transitions}

\begin{figure}[h!]
\centering
\includegraphics[trim=2cm 2cm 4cm 1cm, clip=true,scale=0.43]{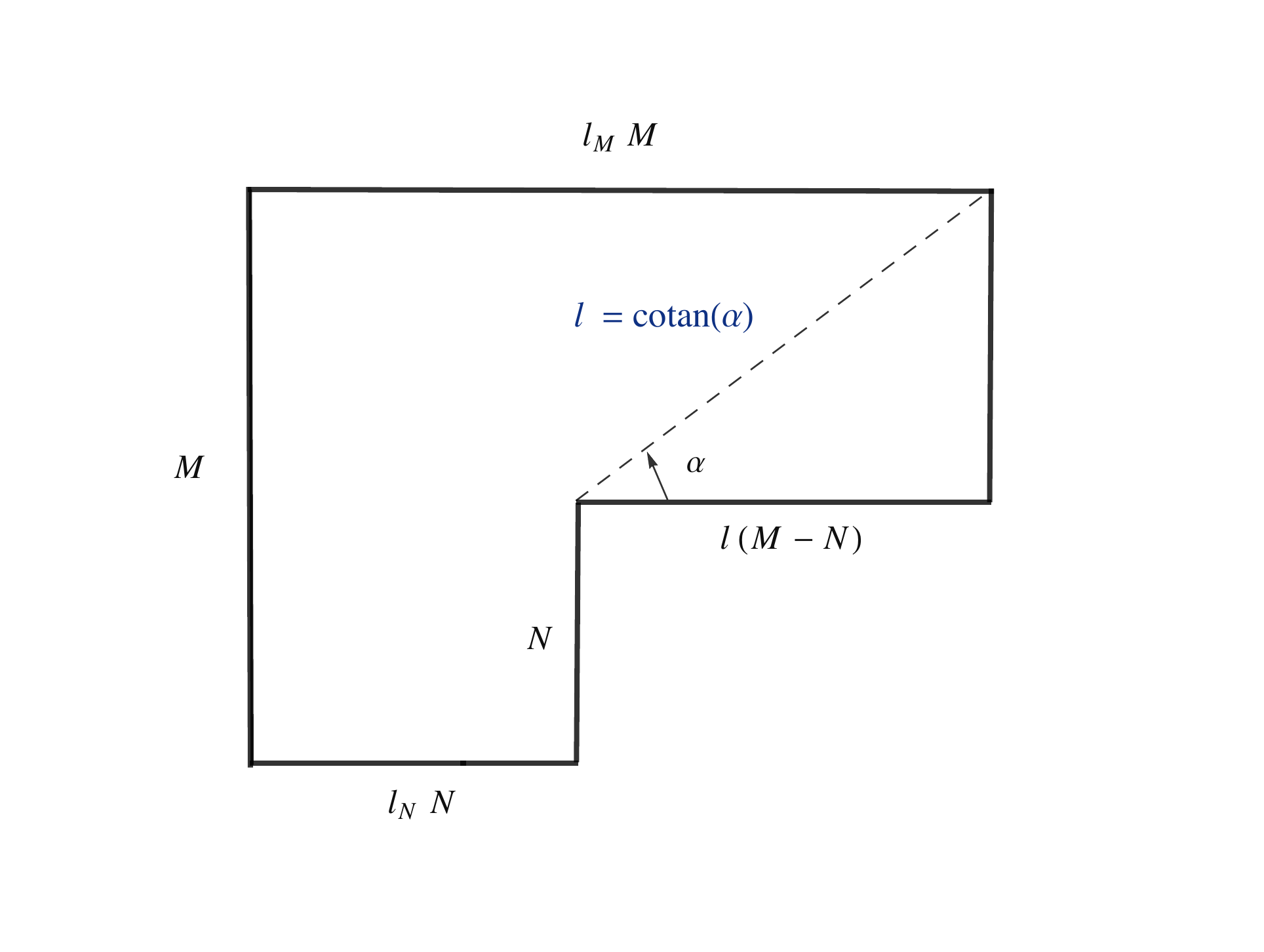}
        \caption{The hook-shaped or two-ring geometry where all the sides have been explicitly written. Parameter $l$
will be relevant for our approaches. It will be properly defined in Eq. (\ref{l_01}). In particular we will consider large $l$, which accounts for thin and long hook shapes.     }\label{cot}
\end{figure}

In this section we compute certain three point functions involving states labeled by Young diagrams $\mu$ and $\nu$ which are stuck at the corners of a ``big'' state $B$, that we will call background and is labeled by a hook-shaped diagram as shown in Fig. \ref{cot}. Later, in section \ref{sec: multi-ring and geometries}, we will associate these $B$ states to LLM geometries. We choose the background state $B$ in such a way that, in a certain limit, the three point functions reproduce the RHS of \eqref{exp}. In the diagram $B$ depicted in Fig. \ref{cot} we will refer to the upper-rightmost corner of this \textit{hook-shaped} diagram as the \textit{$M$-corner}, and to the inward pointing corner as the \textit{$N$-corner}.

\begin{figure}[h!]
\centering
\includegraphics[trim=0cm 1.5cm 0cm 1.6cm, clip=true,scale=0.5]{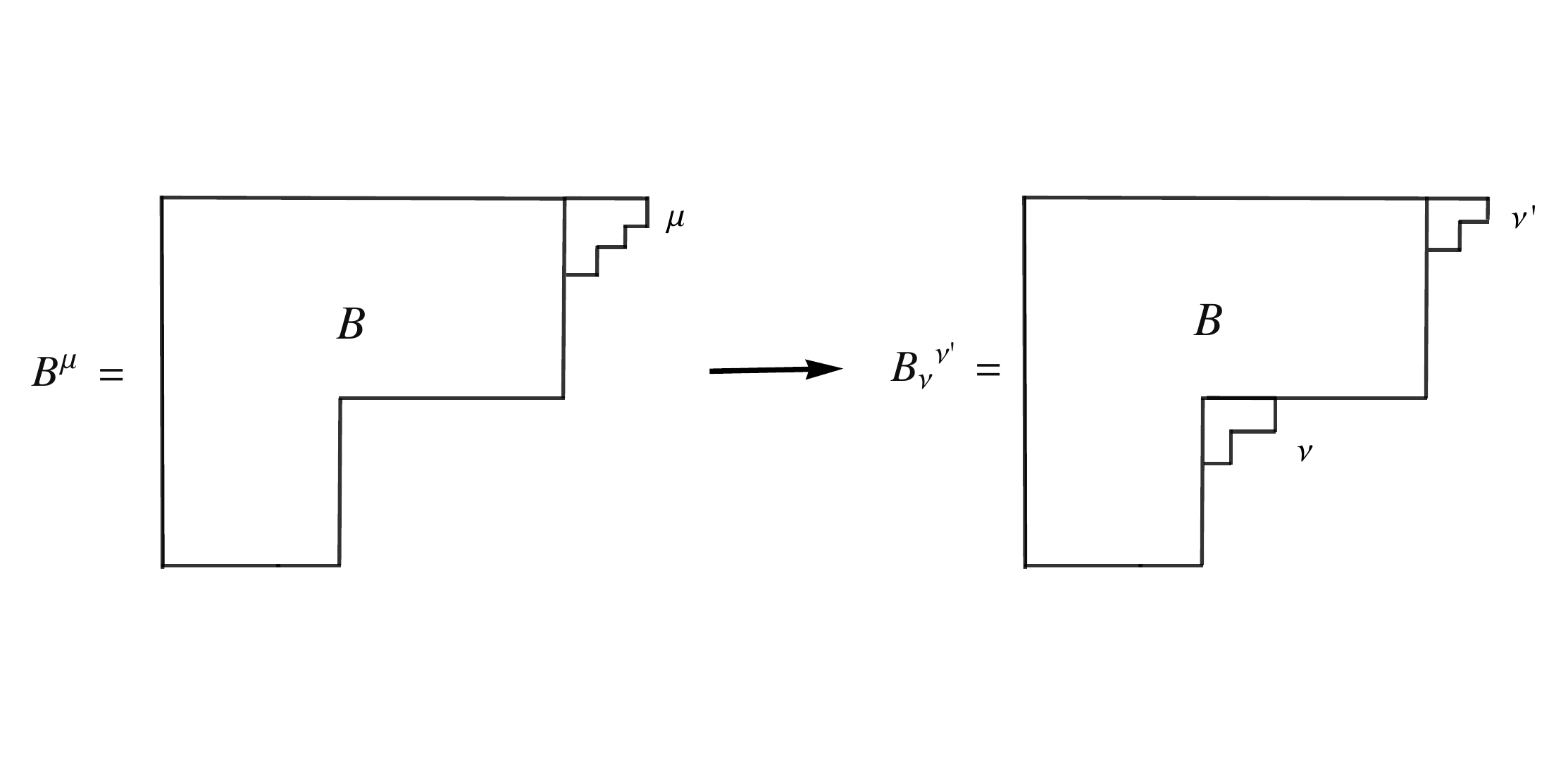}
        \caption{The process whose probabilities are given in Eq. (\ref{prob}). The state $\mu$ in the $M$-corner of the background turns into two states $\nu', \nu$ allocated in the two corners that the background allows. Note that the number of boxes in this process is conserved: $|\mu|=|\nu'|+|\nu|$.   }\label{Transition}
\end{figure}

Concretely, the state corresponding to the above diagram can be produced by acting on the vacuum with a Schur polynomial \cite{Corley:2001zk}
\begin{equation}\label{schurprod}
\chi _{B}(Z)|0\rangle =|B\rangle \,,
\end{equation}
built out of the scalar field $Z$ of the ${\cal N}=4$ gauge theory. By adjoining a Young diagram $\mu$ or $\nu$ either to the $M$-corner or to the $N$-corner, we create excited states of $B$. Since we are considering a $U(M)$ ${\cal N}=4$  theory, we notice that these two corners are the only places where boxes can be attached. The three point functions we compute correspond to processes depicted in Fig. \ref{Transition}, where the number of boxes is conserved, namely
\begin{equation}
|\mu |=m,\quad |\nu |=n,\quad |\nu ^{\prime }|=m-n\,.
\end{equation}%
In the following we will denote
\begin{equation}
P_{\nu }^{\mu \nu
^{\prime }}={\cal P}(B^{\mu }\rightarrow B_{\nu }^{\nu ^{\prime }}),\label{MG}
\end{equation}
the probability of this transition to occur.

In the present work, the backgrounds $B$ that we use to compute the amplitudes are half-BPS, since they are built on a single scalar field $Z$ of the theory\footnote{Although we do not prove it in this work, we claim that the same results would follow if the background state $B$ was constructed using different fields, in which case $B$ could be obtained by the action of restricted Schur operators on the vacuum instead of (\ref{schurprod}). Restricted Schur polynomials are labeled by a big diagram $B$ and smaller ones $b_1,b_2,...$, where the number of small diagrams depends on the number of different fields that are involved in the operator. The relevant diagram would be $B$.}. It is essential in our calculation that the excitations $\mu$ are built on a different field from the background, say $Y$. Had we chosen the same field $Z$ for the excitation, its interaction with the background state would have involved Wick contractions between the background fields and excitation fields, a situation that we want to avoid here. Later in section \ref{sec: multi-ring and geometries}, when we interpret the background state as a geometry and the excitation $\mu$ as a multi-graviton excitation on it, we will think of this condition as saying: we impose that the interaction between $B$ and $\mu$ be purely gravitational. As mentioned before, the excitation states are also half-BPS and as such they are given by Schur polynomials $\chi _{\mu }(Y)$ and $\chi _{\nu }(Y)$, where $\mu $ and $\nu $ are Young diagrams with $m$ and $n$ boxes, respectively. The product of background and excitation can be
written in terms of restricted Schur polynomials as \cite{Bhattacharyya:2008xy}
\begin{equation}
\chi _{B}(Z)\chi _{\mu }(Y)=H_{B}H_{\mu }\sum_{\nu,
\nu ^{\prime },i}\frac{1}{H_{B_{\nu }^{\nu ^{\prime }}}}\chi
_{B_{\nu }^{\nu ^{\prime }},(B,\mu )^{i}}(Z,Y),  \label{partonsproduct}
\end{equation}
where the $B_{\nu }^{\nu ^{\prime }}$ are diagrams that can be formed from
the product $B\times \mu $, and $i$ runs over the multiplicities which are
given by the Littlewood-Richardson numbers $g(B_{\nu }^{\nu ^{\prime}};B,\mu )$. Notice that the operators corresponding to the combination of background and excitation, as in \eqref{partonsproduct}, are quarter-BPS instead of half-BPS.

Let us define the transition probabilities $P_{\nu }^{\mu \nu^{\prime }}$ in terms of correlators of Schur polynomials as
\begin{equation}
P_{\nu }^{\mu \nu
^{\prime }}=\frac{|\langle \chi
_{B}^{\dagger }(Z)\chi _{\mu }^{\dagger }(Y)\chi _{B_{\nu }^{\nu ^{\prime
}},(B,\mu )}(Z,Y)\rangle |^{2}}{\Vert \chi _{B}(Z)\chi _{\mu }(Y)\Vert
^{2}\Vert \chi _{B_{\nu }^{\nu ^{\prime }},(B,\mu )}(Z,Y)\Vert ^{2}}.\label{prob}
\end{equation}
We call them ``probabilities'' because they sum up to 1, that is,
\begin{equation}
\sum_{\nu,\nu ^{\prime }}P^{\mu\nu'}_{\nu}=1,\quad \quad \forall \mu,
\end{equation}
a fact that follows from the observation that (\ref{prob}) is the square of a normalized projection and in (\ref{partonsproduct}) it is shown that all decomposition possibilities of the product $\chi _{B}(Z)\chi _{\mu }(Y)$ into restricted Schur polynomials are exhausted.
Observe that unlike the RHS of equation \eqref{sim}, which is the quantity that we wish to reproduce, $P_{\nu }^{\mu \nu^{\prime }}$ has an explicit dependence on the intermediate states $\nu'$. Hence, we consider instead the trace of $P_{\nu }^{\mu \nu^{\prime }}$ over all the intermediate states and we define $P_{\nu }^{\mu}\equiv{\cal P}(B^{\mu }\rightarrow B_{\nu })$. Explicitly, this transition probability is given by
\begin{equation}
P_{\nu }^{\mu}=\sum_{\nu ^{\prime }}\frac{|\langle
\chi _{B}^{\dagger }(Z)\chi _{\mu }^{\dagger }(Y)\chi _{B_{\nu }^{\nu
^{\prime }},(B,\mu )}(Z,Y)\rangle |^{2}}{\Vert \chi _{B}(Z)\chi _{\mu
}(Y)\Vert ^{2}\Vert \chi _{B_{\nu }^{\nu ^{\prime }},(B,\mu )}(Z,Y)\Vert ^{2}%
}.  \label{probabilities}
\end{equation}
This is the quantity that we compute in the following.

Remember that the two-point function of restricted Schur operators is given by \cite%
{Bhattacharyya:2008rb}
\begin{equation}
\langle \chi _{R,(r,s)}^{\dagger }(Z,Y)\chi _{T,(t,u)}(Z,Y)\rangle =\delta
_{RT}\delta _{rt}\delta _{su}\frac{H_{R}}{H_{r}H_{s}}f_{R}.  \label{restricted}
\end{equation}%
Now, using (\ref{partonsproduct}) and (\ref{restricted}) it is
straightforward to compute the quantities appearing in (\ref%
{probabilities}),
\begin{eqnarray}
\Vert \chi _{B}(Z)\chi _{\mu }(Y)\Vert ^{2} &=&\langle \chi _{B}^{\dagger
}(Z)\chi _{\mu }^{\dagger }(Y)\chi _{B}(Z)\chi _{\mu }(Y)\rangle
=f_{B}f_{\mu },  \notag \\
\Vert \chi _{B_{\nu }^{\nu ^{\prime }},(B,\mu )}(Z,Y)\Vert ^{2} &=&\langle
\chi _{B_{\nu }^{\nu ^{\prime }},(B,\mu )}^{\dagger }(Z,Y)\chi _{B_{\nu
}^{\nu ^{\prime }},(B,\mu )}(Z,Y)\rangle =\frac{H_{B_{\nu }^{\nu
^{\prime }}}}{H_{B}H_{\mu }}f_{B_{\nu }^{\nu ^{\prime
}}},  \notag \\
|\langle \chi _{B}^{\dagger }(Z)\chi _{\mu }^{\dagger }(Y)\chi _{B_{\nu
}^{\nu ^{\prime }},(B,\mu )}(Z,Y)\rangle |^{2} &=&f_{B_{\nu }^{\nu ^{\prime
}}}^{2}g(\mu ;\nu ,\nu ^{\prime }).
\end{eqnarray}%
Above $f_{B}$, $f_{\mu }$, and $f_{B_{\nu }^{\nu ^{\prime}}}$ stand for the \textit{weights} of the Young diagrams $B$, $\mu$, and $B_{\nu }^{\nu ^{\prime}}$ respectively. Note that we have used the Littlewood-Richardson number in the last
equation. It can be shown that for $B$ as in Fig. \ref{Young bubbles} we have
\begin{equation}
g(B_{\nu }^{\nu ^{\prime
}};B,\mu )=g(\mu ;\nu ,\nu ^{\prime }). \label{gB}
\end{equation}
Plugging all these into (\ref%
{probabilities}) we find the expression
\begin{equation}
P_{\nu }^{\mu}=\sum_{\nu ^{\prime }\vdash m-n}g(\mu ;\nu
,\nu ^{\prime })\frac{f_{B_{\nu }^{\nu ^{\prime }}}}{f_{B}f_{\mu }}\frac{%
H_{B}H_{\mu }}{H_{B_{\nu }^{\nu ^{\prime }}}\label{exact Pmn}
}.
\end{equation}%

Observe that the result in Eq. \eqref{exact Pmn} is exact, it is only now that we consider some pertinent approximations. First, let us take the limit of $%
N/M\rightarrow r/r^{\prime }$.
The length of the first row of the Young diagram is
\begin{equation}
l_M M=l_{N}N+l (M-N),\label{l_01}
\end{equation}
and since $N/M\rightarrow r/r^{\prime }$, this means that
\begin{equation}
l_M=l_{N}\left(\frac{r}{r^{\prime }}\right)+l \left(1-\frac{r}{r^{\prime }}\right). \label{l}
\end{equation}
 The weight of the box at the
right-most outward corner is $(l_M+1)M$, while the weight of the box at the
inward corner is $(l_{N}+1)N$. We will also take $n,m\ll N$. In this case we
find\begin{equation}
\frac{f_{B_{\nu }^{\nu ^{\prime }}}}{f_{B}f_{\mu }}\approx \left(\frac{r}{r^{\prime
}}\right)^{n}\Big(\frac{1+l_{N}}{1+l_M}\Big)^{n}(1+l_M)^{m}  \label{fap}
\end{equation}%
for the weights.
For the hooks, it is very important which kind of shape we have chosen for $B$.
Taking into account only the row and column of boxes where the hooks differ, we find
\begin{eqnarray}
H_{B}\times \frac{1}{H_{B_{\Box }}} &=&\frac{%
(N(1+l_{N})-1)!}{(N-1)!}\frac{(M(1+l_M)-N(1+l_{N})-1)!}{(l_M M-l_{N}N-1)!}  \notag
\\
&\times &\frac{(N)!}{(N(1+l_{N}))!}\frac{(l_M M-l_{N}N)!}{(M(1+l_M)-N(1+l_{N}))!}
\notag \\
&=&\frac{1}{1+l_{N}}~\frac{l_M-l_{N}~r/r^{\prime }}{1+l_M-(1+l_{N})r/r^{\prime }}
\notag \\
&=&\frac{1}{1+l_N}\Big[1-\frac{1-r/r^{\prime }}{1+l_M-(1+l_{N})r/r^{\prime }}%
\Big].
\end{eqnarray}%
In general, for $|\nu |=n$, we will have
\begin{equation}
H_{B}\times \frac{1}{H_{B_{\nu }}}\approx \Big(\frac{1%
}{1+l_{N}}\Big)^{n}\Big[1-\frac{1-r/r^{\prime }}{1+l_M-(1+l_{N})r/r^{\prime }}%
\Big]^{n}\frac{1}{H_{\nu }}.  \label{up}
\end{equation}%
Analogously, for boxes in the upper corner of $B$ we find
\begin{eqnarray}
H_{B}\times \frac{1}{H_{B^{\Box }}} &=&\frac{%
(M(1+l_M)-1)!}{(M(1+l_M)-l_{N}N-1)!}~\frac{(M(1+l_M)-N(1+l_{N})-1)!}{(M-N-1)!}
\notag \\
&\times &\frac{(M(1+l_M)-l_{N}N)!}{(M(1+l_M))!}~\frac{(M-N)!}{%
(M(1+l_M)-N(1+l_{N}))!}  \notag \\
&=&\frac{1+l_M-l_{N}~r/r^{\prime }}{1+l_M}~\frac{1-r/r^{\prime }}{%
1+l_M-(1+l_{N})r/r^{\prime }}  \notag \\
&=&\frac{1-r/r^{\prime }}{1+l_M}\Big[1+\frac{r/r^{\prime }}{%
1+l_M-(1+l_{N})r/r^{\prime }}\Big].
\end{eqnarray}%
Thus, for states $|\nu ^{\prime }|=m-n$, we will have
\begin{equation}
H_{B}\times \frac{1}{H_{B^{\nu ^{\prime }}}}\approx %
\Big(\frac{1-r/r^{\prime }}{1+l_M}\Big)^{m-n}\Big[1+\frac{r/r^{\prime }}{%
1+l_M-(1+l_{N})r/r^{\prime }}\Big]^{m-n}\frac{1}{H_{\nu ^{\prime }}}%
.  \label{down}
\end{equation}%
Combining (\ref{up}), (\ref{down}) and (\ref{fap}), we can write
every summand of the final probability as
\begin{eqnarray}
\frac{f_{B_{\nu }^{\nu ^{\prime }}}}{f_{B}f_{\mu }}\frac{H_{B}%
H_{\mu }}{H_{B_{\nu }^{\nu ^{\prime }}}}
&=&\left(\frac{r}{r^{\prime }}\right)^{n}\left(1-\frac{r}{r^{\prime }}\right)^{m-n}  \notag  \label{summands} \\
&\times &\Big[1-\frac{1-r/r^{\prime }}{l_M+1-(l_{N}+1)r/r^{\prime }}\Big]^{n}%
\Big[1+\frac{r/r^{\prime }}{l_M+1-(l_{N}+1)r/r^{\prime }}\Big]^{m-n}  \notag \\
&\times &\frac{H_{\mu }}{H_{\nu ^{\prime }}H_{\nu }}.
\end{eqnarray}%
Recall that the hooks relate to the dimensions of irreps of the
symmetric group via
\begin{equation}
H_{R}=\frac{n!}{\text{dim}_{R}},
\end{equation}%
which implies
\begin{equation}
\frac{H_{\mu }}{H_{\nu ^{\prime }}H_{\nu }}=%
\frac{m!}{(m-n)!n!}\frac{\text{dim}_{\nu ^{\prime }}\text{dim}_{\nu }}{\text{%
dim}_{\mu }}.
\end{equation}%
 So we have
\begin{eqnarray}
P_{\nu }^{\mu\nu'} &=&\left(\frac{r}{r^{\prime }}\right)^{n}\left(1-\frac{r}{r^{\prime }}\right)^{m-n}
\notag \\
&\times &\Big[1-\frac{1-r/r^{\prime }}{l_M+1-(l_{N}+1)r/r^{\prime }}\Big]^{n}%
\Big[1+\frac{r/r^{\prime }}{l_M+1-(l_{N}+1)r/r^{\prime }}\Big]^{m-n}  \notag \\
&\times &\frac{m!}{(m-n)!n!} \;  g(\mu ;\nu ,\nu ^{\prime })\;\frac{\text{dim}_{\nu ^{\prime
}}\text{dim}_{\nu }}{%
\text{dim}_{\mu }}.\label{Pmnn}
\end{eqnarray}%

Finally, using Eq. \eqref{l} as well as
\begin{equation}
\sum_{\nu ^{\prime }}g(\mu ;\nu ,\nu ^{\prime })\,\text{dim}_{\nu ^{\prime
}}= \text{ dim}(\mu ,\nu ),
\end{equation}%
we obtain
\begin{equation}
P_{\nu }^{\mu}=\left( \frac{r}{r^{\prime }}\frac{l}{%
l+1}\right) ^{n}\left( 1-\frac{r}{r^{\prime }}\frac{l}{l+1}%
\right) ^{m-n}\frac{m!}{(m-n)!n!}\frac{\text{ dim}(\mu ,\nu )\text{dim}_{\nu
}}{\text{dim}_{\mu }}.  \label{final2}
\end{equation}%
Notice that the above expression matches exactly the RHS of (\ref{sim}) in the limit
$l\rightarrow\infty$. The outcome is that the Young-Bouquet probability distribution \eqref{YB dist} is produced by the large $l$ limit\footnote{This limit has a picture for LLM geometries as discussed  in section \ref{sec: multi-ring and geometries}. We will see that in the LLM geometries, this regime corresponds to two well separated black rings in Fig. \ref{Young bubbles}.} of the transition probability $P_{\nu }^{\mu}$. This is the main result of this section.

We would like to end this section with a couple of comments. First, the shape of the background $B$ that we have chosen
has been crucial for the computation to hold: $\mu$ and $\nu$ could be added only at the two corners of $B$. This translates into the decomposition (\ref{partonsproduct}) and the probabilistic interpretation of (\ref{prob}) on the one hand, and also leads to the non-trivial relation (\ref{gB}). Second, we would like to comment on the approximations that we
have made. It was shown in \cite{BO} that (\ref{sim}) is exact in the limit $%
N,M\rightarrow \infty $ but would have $1/N$ corrections for finite $N$. We
have also suppressed $1/N$ corrections, which are of order $n/N$, when
computing the quotient of Hooks and functions $f$ in (\ref{fap}), (\ref{up})
and (\ref{down}). This is important because we could
allow the excitations to be composite operators that grow with $N$ in a certain way.
For example, excitations of strings which are driven by
operators of length $n\sim \sqrt{N}$, are expected to behave well, so the computation
would still hold for those excitations. However,
with Giant Gravitons, whose dual operators grow as $n\sim N$, the approximation
would break down. It seems that the approximation holds for perturbative
objects.

\section{The eigenvalues of the embedding chain charges}

\label{sec: conserved charges}

In the previous section we have shown that the probabilities in the RHS of (\ref{sim}) can be obtained by considering a particular physical process. Something similar can be done for the quantities appearing in the LHS of (\ref{sim}). These quantities, the $\mathbb{GT}$ distributions, are intimately related
to the eigenvalues of the ``embedding'' observables found in \cite{Diaz:2013gja,Diaz:2014ixa}.
In those works, the infinite embedding
chain\footnote{The link of the embedding chain (\ref{ec}) and representation theory is clear: choose an irrep of $U(M)$ labeled by the diagram $\mu$, now any $M-1$ collection of irreps belonging to the groups $U(M-1),\dots, U(1)$ such that the irrep of $U(k-1)$ is subduced by the irrep of $U(k)$ is a state of irrep $\mu$. This is the idea that lies behind the construction of Gelfand-Tsetlin patterns.} of Lie \textit{algebras}
\begin{equation}\label{ec}
\mathfrak{u}(1)\hookrightarrow \mathfrak{u}(2)\hookrightarrow \cdots
\hookrightarrow \mathfrak{u}(N)\hookrightarrow \cdots
\end{equation}%
was considered and a set of conserved observables, that we denominate charges, in the free $U(N)$ CFT consistent with the chain was found.

Let us comment on the name ``charges'' for those observables. Charges are usually constructed from currents and the latter are built from elementary fields of the theory. This is not the way that the observables in \cite{Diaz:2013gja,Diaz:2014ixa} were constructed. In this sense, the observables that we call charges are not conventional charges since they don't come directly from a symmetry of the theory. The operators were constructed by embedding states of the $U(N)$ CFT into a $U(M)$ CFT and projecting them back to the $U(N)$ theory in a unique and consistent way\footnote{Although it is necessary to choose a particular embedding in the process, the observables are embedding independent in the final result.}. So, they act on states and give back states of the $U(N)$ CFT. We call them charges for two reasons: first, they are conserved since they commute with the Wick contractions which carry all the dynamics of the free theory, and second, they resolve the label of the Schur polynomial states, which we understand as quantum labels\footnote{This point of view in which these embedding observables resolve quantum labels gets enhanced in \cite{Diaz:2014ixa}, where analogous embedding observables are seen to resolve the small labels of restricted Schur polynomials.}.

The result of \cite{Diaz:2013gja,Diaz:2014ixa} was an infinite tower of charges $\{Q_{NM}|~M>N\}$
which, on half-BPS states, behave as\footnote{As studied in \cite{Diaz:2014ixa} they act in all the states of the spectrum, not just half-BPS.}
\begin{equation}
Q_{NM}\chi _{\mu }(Y)=\frac{f_{\mu }(N)}{f_{\mu }(M)}\chi _{\mu }(Y),\label{charges}
\end{equation}%
so, their eigenvectors are Schur polynomials and the eigenvalues are given by
\begin{equation}
\frac{f_{\mu }(N)}{f_{\mu }(M)}=\frac{\text{Dim}[\mu ,N]}{\text{Dim}[\mu ,M]}\,.
\end{equation}%
It is easy to see that there is just one path in $\mathbb{GT}$ joining
$[\mu ,M]$ with $[\mu ,N]$, that is,
\begin{equation}
\text{Dim}([\mu ,M],[\mu ,N])=1.
\end{equation}%
So, in fact,
\begin{equation}\label{chid}
\frac{f_{\mu }(N)}{f_{\mu }(M)}=\frac{\text{Dim}[\mu ,N]~\text{Dim}([\mu
,M],[\mu ,N])}{\text{Dim}[\mu ,M]}={}^{\mathbb{GT}}\Lambda _{N}^{M}(\mu ,\mu
).
\end{equation}%
The eigenvalues of the charges \eqref{charges} are actually probabilities of $\mathbb{GT}$, as we can see. They apply to arbitrary (also
infinite) $N$ and $M$, as long as $M>N$. Of course one can take the limit $%
N,M\rightarrow \infty $ with $N/M\rightarrow r/r^{\prime }$ and reproduce
the LHS of (\ref{sim}) for the cases that the two states are labeled by the
same Young Diagram.

The link between the eigenvalues of $Q_{\mu\nu}$ and the probabilities of the ${\mathbb{GT}}$ graph is not surprising since both are constructed from the embedding chain (\ref{ec}). What is not obvious at all is that these probabilities, or the eigenvalues of the charges, have anything to do with the three point functions computed in section \ref{sec: multi-graviton transitions}, that is, with the RHS of (\ref{sim}). Thus, the BO identity makes a natural association between two apparently different objects: the eigenvalues of the embedding chain charges and the three point functions on hook-shaped backgrounds. It is precisely this surprising match, and the fact that the background has a natural gravity identification through bubbling geometries, that can be used in order to give a holographic interpretation of the embedding charges. We will briefly discuss these in subsection \ref{holointer}.

Two comments are in order. First, since the charges were constructed for arbitrary $N$, they are
observables with physical meaning for finite $N$. This suggests that the
meaningful extension of (\ref{sim}) for finite $N$ should be found by
keeping the LHS and modifying the RHS. This can be done by taking into account
$1/N$ corrections in the preceding section, in which the RHS of (\ref{sim}) was
reproduced by the three point functions. Studying finite $N$ corrections
of (\ref{sim}) seems an interesting mathematical and physical question.

Second, we could consider the embedding chain of different algebras. In \cite%
{Diaz:2013gja,Diaz:2014ixa} the orthogonal and symplectic algebras were also
considered. The treatment is similar to the unitary case with the obvious
technical differences. Analogous embedding charges were found. However, in
the case of orthogonal and symplectic algebras the relevant finite groups
were not ordinary symmetric groups but \textit{wreath products} $S_{n}[S_{2}]$. This
suggests that an analogous identity to (\ref{sim}) for orthogonal and
symplectic gauge groups might exist, where the graph of the RHS is the one
of wreath products.

\section{Background states and multi-ring geometries}

\label{sec: multi-ring and geometries}

In section \ref{sec: multi-graviton transitions} we see that the distributions associated with the $\mathbb{YB}/\mathbb{GT}$ duality can be encoded in certain transition probabilities computed in ${\cal N}=4$ gauge theory. It is crucial though that those correlators involve certain background states: hook-shaped backgrounds. We recall that these backgrounds are also half-BPS since they are built on a single scalar of the theory. For large $N$, the background states are very heavy and, as shown in \cite{Lin:2004nb,Corley:2001zk,Berenstein:2004kk,Koch:2008ah}, there is a correspondence between those heavy states and bubbling geometries. As we will see below, they correspond to two concentric rings in the bubbling plane.  We have already hinted at the procedure that must be followed to achieve the description in Fig. \ref{Young bubbles}. The purpose of this section is to perform this task explicitly. Studying the IR and the UV limits of the two-ring geometry, we will show that a hook-shaped background state corresponds to a geometry that interpolates between two AdS spaces with different radii. Besides, we will see that the same computation that we performed in section \ref{sec: multi-graviton transitions} would follow if the background has a multi-ring structure under certain conditions. This resembles the compatibility condition \eqref{compa}. Actually, we will show that the compatibility condition \eqref{compa} can be understood in terms of ring splitting in the bubbling plane.

\subsection{Multi-ring geometries}

\label{sec: multi-ring}

Let us briefly remind the reader the physical meaning behind the construction of these\textit{ bubbling geometries}.
Half-BPS states in ${\cal N}=4$ gauge theory can be labeled by Young diagrams, and to each of these diagrams we can associate a black and white pattern in the plane. Finally, provided that the diagram is large enough, this pattern provides the necessary boundary conditions to reconstruct the full ten dimensional geometry coming from string theory \cite{Corley:2001zk,Berenstein:2004kk,Lin:2004nb}. The plane containing such patterns is called the bubbling plane. The geometries obtained by this procedure correspond to pure states of the theory, and their coarse graining has been discussed in  \cite%
{Balasubramanian:2005mg,Bellucci:2009pb}. The details of these microstate geometries can be distinguished by Young diagrams and their correlation
functions in the field theory, see for example \cite{Koch:2008ah,Skenderis:2007yb,Lin:2010sba,Ghodsi:2005ks}. The excitations of these geometries are described by the UV complete string theory. The physical properties of microstate geometries in various other situations have been discussed extensively \cite{Bena:2006is,Chowdhury:2007jx}.

\begin{figure}
\centering
\includegraphics[scale=0.38]{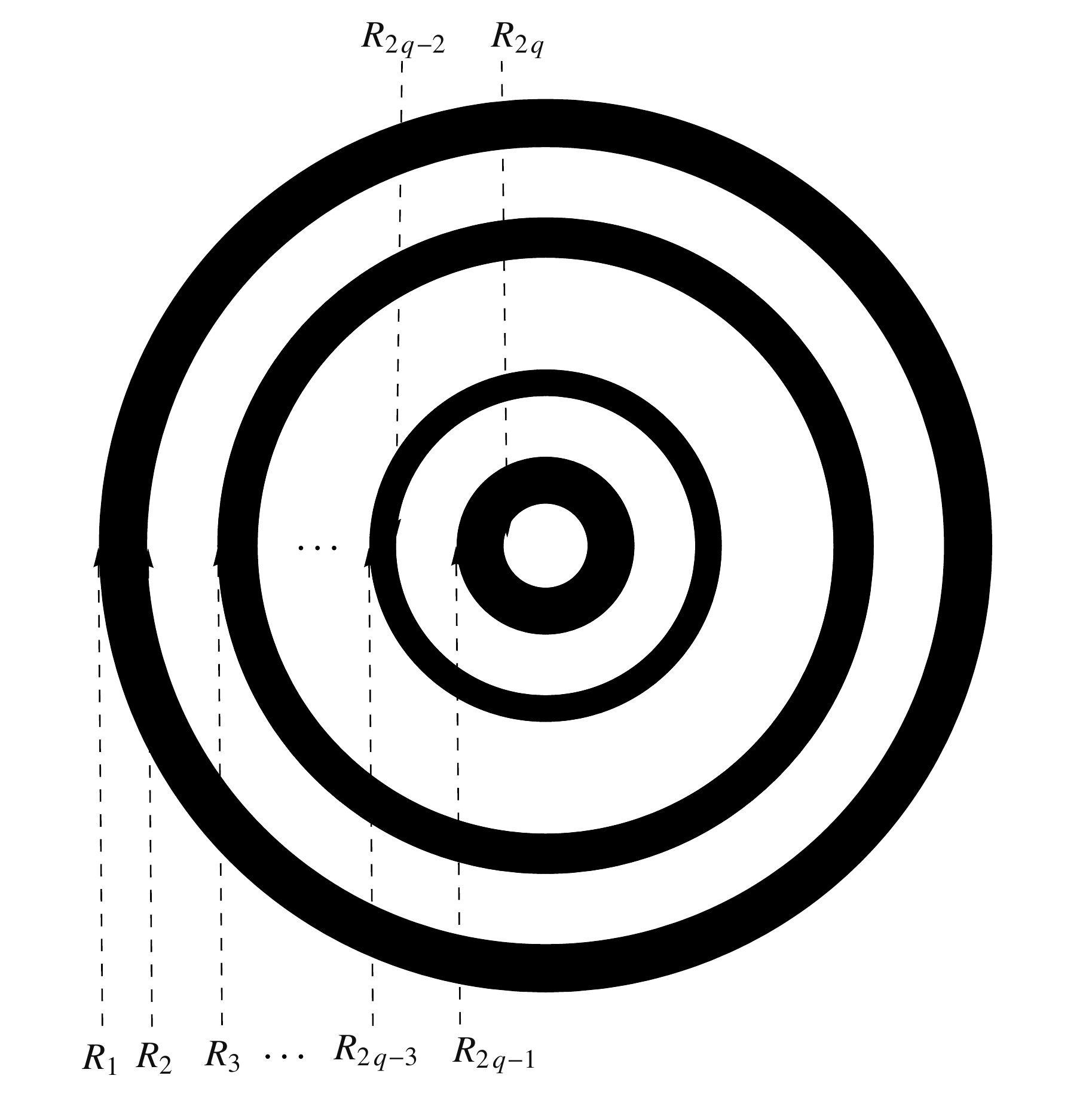}
        \caption{Multi-ring structures in the bubbling plane.}\label{multiring}
\end{figure}

Let us now consider a pattern of droplets given by $q$ concentric black rings, that is, with $2q$ circles (see Fig. \ref{multiring}). The line element corresponding to these geometries can be written as \cite{Lin:2004nb}
\begin{eqnarray}
ds^{2} &=&-h^{-2}(dt+Vd\phi )^{2}+h^{2}(dy^{2}+dr^{2}+r^{2}d\phi
^{2})+ye^{G}d\Omega _{3}^{2}+ye^{-G}d\tilde{\Omega}_{3}^{2},
\label{metric_01} \\
h^{-2} &=&\frac{2y}{\sqrt{1-4z^{2}}},~~~~~e^{G}=\sqrt{\frac{1+2z}{1-2z}}%
,~~~~re^{i\phi }=x_{1}+ix_{2}.
\end{eqnarray}%
The $t$ is the time coordinate, and $y$ and $r$ are two radial-like coordinates. The coordinates ($x_{1},x_{2}$) can also be viewed as a phase space plane.  The functions $z$ and $V$ in the above expressions are given by
\begin{eqnarray}
z &=&\frac{1}{2}+\tilde{z}=\frac{1}{2}+\sum_{i=1}^{2q}(-1)^{i+1}\left(
\frac{r^{2}+y^{2}-R_{i}^{2}}{2\sqrt{%
(r^{2}+y^{2}+R_{i}^{2})^{2}-4r^{2}R_{i}^{2}}}-\frac{1}{2}\right)\nonumber \\
&=&\frac{1}{2}+\sum_{i=1}^{2q}(-1)^{i+1}\tilde{z}(R_{i}^{2}), \\
V &=&-\sum_{i=1}^{2q}(-1)^{i+1}\left( \frac{r^{2}+y^{2}+R_{i}^{2}}{2\sqrt{%
(r^{2}+y^{2}+R_{i}^{2})^{2}-4r^{2}R_{i}^{2}}}-\frac{1}{2}\right)
\label{V_01}\nonumber  \\
&=&\sum_{i=1}^{2q}(-1)^{i+1}V(R_{i}^{2}).  \label{V_02}
\end{eqnarray}%
For future convenience, we define the functions%
\begin{eqnarray}
\tilde{z}(R_{i}^{2}) &=&\left( \frac{r^{2}+y^{2}-R_{i}^{2}}{2\sqrt{%
(r^{2}+y^{2}+R_{i}^{2})^{2}-4r^{2}R_{i}^{2}}}-\frac{1}{2}\right) ,
\label{tilde_z_i_01} \nonumber\\
V(R_{i}^{2}) &=&-\left( \frac{r^{2}+y^{2}+R_{i}^{2}}{2\sqrt{%
(r^{2}+y^{2}+R_{i}^{2})^{2}-4r^{2}R_{i}^{2}}}-\frac{1}{2}\right) .
\label{V_i_01}
\end{eqnarray}
Due to the regularity condition, the function $z$ at the $y=0$ plane, takes values of $-1/2$ and $1/2$ in two different regions of the phase space plane, which are denoted as black and white regions, respectively. In the expressions above, the $R_{i}$'s stand for the radii of the $2q$ rings that form the $q$ black rings in Fig. \ref{multiring}. These radii are sequenced in the radial direction of the phase space plane
\begin{equation}\label{radii}
R_{2q}<...<R_{i+1}<R_{i}<...<R_{1}.
\end{equation}%
The rings correspond to topological cycles of the geometries.

The flux quantization requires the areas of the droplets to be quantized.
We refer to the area of the $j$-th black droplet as (Area)$_{j}^{b}$, while $%
\tilde{N}_{j}$ stands for the quanta of fluxes through that droplet. For the black
droplets we have
\begin{equation}
\frac{(\text{Area})_{j}^{b}}{4\pi ^{2}l_{p}^{4}}=\tilde{N}%
_{j}=R_{2j-1}^{2}-R_{2j}^{2},
\end{equation}%
where in the last equation we have chosen an appropriate unit for the area. On the other hand,
for the white droplets we find
\begin{equation}
\frac{(\text{Area})_{j}^{w}}{4\pi ^{2}l_{p}^{4}}=l_{j}\tilde{N}%
_{j}=R_{2j}^{2}-R_{2j+1}^{2},
\end{equation}%
where the $l_{j}\tilde{N}_{j}$ are the flux quanta threading the white
droplets.

Let us relate these facts to a Young diagram and take the first row of the diagram in question
to be of length $l_M M$, then summing over the $q$ black rings we have
\begin{equation}
l_M M=\sum_{j=1}^{q}l_{j}\tilde{N}_{j}.
\end{equation}
If we define
\begin{equation}
N_{j}=\sum_{j^{\prime }\geqslant j}\tilde{N}_{j^{\prime }},
\end{equation}%
 we get a sequence of increasing integers $\{N_{j}|~j=1,...,q\}$:%
\begin{equation}\label{integersequence}
N=N_{q}<...<N_{j+1}<N_{j}<...<N_{1}=M.
\end{equation}%
This sequence of integers can be mapped to the Gelfand-Tsetlin chain of embeddings
\begin{equation}
U(M)\supset ...\supset U(N_{j})\supset U(N_{j+1})\supset ...\supset U(N).
\label{GT_embedding_01}
\end{equation}%
With this identification, the further we go towards the center of the phase space plane, the less rings
we explore and the smaller the gauge groups in the embedding (\ref{GT_embedding_01}).

Now, we consider the Gelfand-Tsetlin embedding
\begin{equation}
U(M)\supset U(N),
\end{equation}%
which, under the identifications of chains (\ref{integersequence}) and (\ref{GT_embedding_01}), corresponds to two-ring geometries which in turn are associated to hook-shaped Young diagrams. In particular, we consider diagrams with edges whose lengths, from the left-most corner to the
right-most corner, are $l_{N}N$, $N$, $l_{M}M-l_N N$ and $M-N$, respectively (see Fig. \ref{cot}). Let's denote the radii of the four circles, from the
outer-most circle to the inner-most circle, to be $R_{1},R_{2},R_{3}$ and $R_{4}$ (see Fig. \ref{Young bubbles}). From the relationship to the flux quantization, the four radii are given by%
\begin{eqnarray}
R_{1} &=&\sqrt{(l_{M}+1)M}~~~~~~~R_{2}=\sqrt{%
l_{M}M+N},~  \notag \\
R_{3} &=&\sqrt{(l_{N}+1)N}~~~~~~~R_{4}=\sqrt{l_{N}N}.  \label{radii_01}
\end{eqnarray}%
Plugging Eq. (\ref{radii_01}) into Eqs. (\ref{metric_01})-(\ref{V_02}) we can find the explicit form of the two-ring geometry.

\subsection{UV and IR limits}

\label{sec: limits}

In the previous subsection we outlined the construction of the geometry corresponding to a two-ring pattern in the bubbling plane.
Here, we consider two limits of this geometry and see that they approach two AdS spaces with different radii. Let us begin this discussion with some geometrical observations. The radial coordinates $(r,y)$ in \eqref{metric_01} relate to the AdS radial direction $\rho$ via  \cite{Lin:2004nb}
\begin{equation}
r=\sqrt{R_{AdS}}\cosh \rho \cos \theta ,~~\ y=\sqrt{R_{AdS}}\sinh \rho \sin
\theta ,~~
\end{equation}
and observe that in the limit $\rho \rightarrow \infty$, we have
\begin{equation}
\frac{y}{r}=\tan\theta ,
\end{equation}
 so $r^{2}+y^{2}$ plays the role of the AdS radial direction $\sinh^{2}\rho$. It will come in handy in the following discussion to recall that in the bubbling plane, the $AdS\times S$ spacetime corresponds to a single black disk of radius
$R_0$ and the functions \eqref{tilde_z_i_01} entering the geometry can be expressed as
\begin{eqnarray}
z &=&\frac{1}{2}+\tilde{z}(R_{0}^{2})=\frac{1}{2}-\frac{y^{2}R_{0}^{2}}{%
(r^{2}+y^{2})^{2}}+{\cal O}\left( \frac{1}{(r^{2}+y^{2})^{4}}\right) ,\nonumber \\
V &=&V(R_{0}^{2})=-\frac{r^{2}R_{0}^{2}}{(r^{2}+y^{2})^{2}}+{\cal O}\left( \frac{1}{%
(r^{2}+y^{2})^{4}}\right) ,\label{AdS S}
\end{eqnarray}
in the large radius limit.

First, let us consider the regime
\begin{equation}\label{limit 1}
R_{3}^{2}\ll r^{2}+y^{2}\ll R_{1}^{2},
\end{equation}%
which lies in between the two black rings and corresponds to the infrared limit of the field
theory. Expanding the two-ring geometry in this limit we find
\begin{eqnarray}
z &=&\frac{1}{2}+\sum_{i=1}^4(-1)^{i+1}\tilde{z}(R_{i}^{2})=\frac{1}{2}-%
\frac{y^{2}(R_{3}^{2}-R_{4}^{2})}{(r^{2}+y^{2})^{2}}+{\cal O}\left( \frac{1}{%
(r^{2}+y^{2})^{4}}\right) , \nonumber \label{z_limit_1} \\
V &=&\sum_{i=1}^4(-1)^{i+1}V(R_{i}^{2})=-\frac{r^{2}(R_{3}^{2}-R_{4}^{2})%
}{(r^{2}+y^{2})^{2}}+{\cal O}\left( \frac{1}{(r^{2}+y^{2})^{4}}\right) .
\label{V_limit_1}
\end{eqnarray}%
Comparing equations \eqref{AdS S} and \eqref{V_limit_1} we see that the above geometry approaches an $AdS\times S$ space with
bubbling plane radius
\begin{equation}
R_0=\sqrt{R_3^2-R_4^2}.
\end{equation}
 Moreover, from equations \eqref{radii_01} we know that
\begin{equation}
R_{3}^{2}-R_{4}^{2}=N.
\end{equation}
Therefore, in the limit \eqref{limit 1} the two-ring geometry approaches an $AdS\times S$ spacetime with radius $R_{AdS}^{4}=4\pi ^{2}Nl_{p}^{4}$.

Now we turn to the regime
\begin{equation}\label{limit 2}
r^{2}+y^{2}\gg R_{1}^{2},
\end{equation}%
which corresponds to the asymptotic region of the spacetime. In this limit, the expansion of the two-ring geometry reads
\begin{eqnarray}
z &=&\frac{1}{2}+\sum_{i=1}^{4}(-1)^{i+1}\tilde{z}(R_{i}^{2})=\frac{1}{2}-%
\frac{y^{2}(R_{1}^{2}+R_{3}^{2}-R_{2}^{2}-R_{4}^{2})}{(r^{2}+y^{2})^{2}}%
+O\left( \frac{1}{(r^{2}+y^{2})^{4}}\right) ,\nonumber  \\
V &=&\sum_{i=1}^{4}(-1)^{i+1}V(R_{i}^{2})=-\frac{%
r^{2}(R_{1}^{2}+R_{3}^{2}-R_{2}^{2}-R_{4}^{2})}{(r^{2}+y^{2})^{2}}+O\left(
\frac{1}{(r^{2}+y^{2})^{4}}\right) .  \label{V_limit_2}
\end{eqnarray}%
Just as before, comparing equations \eqref{AdS S} and \eqref{V_limit_2} we notice this geometry approaches an
$AdS\times S$ space, but now the bubbling plane radius is given by
\begin{equation}
R_0=\sqrt{R_{1}^{2}+R_{3}^{2}-R_{2}^{2}-R_{4}^{2}}.
\end{equation}
Furthermore, equations \eqref{radii_01} imply
\begin{equation}
 R_{1}^{2}+R_{3}^{2}-R_{2}^{2}-R_{4}^{2}=M.
\end{equation}
Hence, in the limit \eqref{limit 2} the two-ring geometry approaches an $AdS\times S$ spacetime as well but this time with an AdS radius $R_{AdS}^{4}=4\pi ^{2}Ml_{p}^{4}$. We conclude that the two-ring geometry interpolates between an $(AdS\times S)_{M}$ near the spacetime infinity and an $(AdS\times S)_{N}$ throat in the interior. These results provide a geometric picture of the background $B$ of Fig. \ref{cot}.

Observe that the results in this subsection are valid also for other backgrounds $B$ corresponding to different (large) Young diagrams having more rings; the only condition is that the inner black rings must be well separated from the outermost one. More precisely, for a sequence of radii like \eqref{radii} this condition is satisfied whenever
\begin{equation}
\frac{R_{2}-R_{3}}{R_{1}-R_{2}}\gg 1.
\end{equation}
In the language of section \ref{sec: multi-ring}, if we take the area of the outermost black ring to be $l(M-N)$ then we have
\begin{equation}
\frac{R_{2}-R_{3}}{R_{1}-R_{2}}= l\;\left [1+{\cal O}\left(\frac{l_{N}}{l}\frac{N}{M}\right)\right].
\end{equation}
This means that the large separation between the inner black rings and outermost one can be guaranteed by taking $l\gg 1$. In terms of Fig. \ref{cot}, this condition means that we must consider Young diagrams where the angle $\alpha$ is small.

\subsection{Excitations of the background and holography}\label{holointer}
We have described the dual gravity picture of the background $B$ in the previous subsections. In this subsection we will go further and propose a gravity dual of the processes described in section \ref{sec: multi-graviton transitions}. The background state $B$ in Fig. \ref{cot} is identified to be dual to the two-ring geometry. Given the correspondence between background states and geometries that we have seen in the previous subsections, it is natural to interpret the small Young diagrams $\mu$, $\nu$ and $\nu'$ in the $M$ and $N$ corners as excitations of this background $B$ \cite{Lin:2004nb,Koch:2008ah,Lin:2010sba}. These Young diagrams attached at the corners of $B$ are much smaller in size than the diagram $B$, hence we call them small Young diagrams here. The small Young diagrams $\mu$ and $\nu'$ in the $M$ corner can be seen to be dual to multi-graviton excitations on the $R_{1}$ circle in the two-ring geometry (see Fig. \ref{Young bubbles}). On the other hand, the small Young diagram $\nu$ in the $N$ corner can be seen to be dual to multi-graviton excitations on the $R_{3}$ circle in the two-ring geometry.

In section \ref{sec: multi-graviton transitions} we have shown that the $\mathbb{YB}$ distribution, which is in the RHS of (\ref{sim}), is produced by computing transition probabilities of the form depicted in Fig. \ref{Transition}. The initial state, as shown in that figure, is the background with a small diagram $\mu$ attached to the top corner of the background, and the final state of the transition is the background with diagrams $\nu$ and $\nu'$ attached to each corner. The dual description of this transition process in the spacetime is a transition process of multi-gravitons in the two-ring geometry in Fig. \ref{Young bubbles}. That is, we claim that the field theory process described in section \ref{sec: multi-graviton transitions} is dual to the spacetime process of multigraviton states propagating on the two-ring geometry that is dual to the background $B$. Following this logic, the transition probability $P_{\nu }^{\mu \nu ^{\prime }}$ discussed in section \ref{sec: multi-graviton transitions} is conjectured to correspond to the probability that a multigraviton excitation with angular momentum $m$ on $(AdS\times S)_{M}$ decays into two multigraviton excitations, one on $(AdS\times S)_{N}$  with angular momentum $n\leq m$ and another one on $(AdS\times S)_{M}$ with angular momentum $m-n$. Analogously, the quantity $P_{\nu }^{\mu}$ is conjectured to correspond to the probability that the multigraviton $\mu$ produced at one region propagates to the multigraviton $\nu$ at the other region in the two-ring geometry that we are considering. We have already identified the spacetime background that the dual transition processes take place in. Certain characteristics of the dual transition processes have also been found above, however, we have not performed a direct spacetime computation of the transition probabilities of these processes in spacetime. It would be very interesting to actually perform this calculation in spacetime and confirm our claim. That calculation seems tough, though, and it is out of the scope of this article.

Renormalization group flows have been described holographically using domain-wall spacetimes \cite{Girardello:1998pd,Bianchi:2001de}. These are geometries that interpolate between two AdS spacetimes $AdS_{UV}$ and $AdS_{IR}$ with radii $R_{UV}>R_{IR}$, such as the two-ring spacetimes discussed above. The AdS/CFT dictionary relates these radii to the ranks of the gauge groups of the respective dual field theories via $R_{UV}^{4}=4\pi ^{2}Ml_{p}^{4}$ and $R_{IR}^{4}=4\pi ^{2}Nl_{p}^{4}$. The claim is that as we traverse the domain wall from $AdS_{UV}$ towards $AdS_{IR}$ we flow from a theory with gauge group $U(M)$ towards one with a $U(N)$ gauge group, thus along these trajectories we would be effectively integrating out the degrees of freedom in $U(M-N)\subset U(M)/U(N)$. This argument seems to suggest that the LHS of (\ref{sim}) which we identify with field theory data, can be seen as the probabilities that state $\mu$ in $U(M)$ field theory descends to state $\nu$ in $U(N)$ field theory through an RG flow.

To analyze our above idea we could start by assuming that the process occurring in the field theory is an RG flow, and ask what is the holographic picture of that process in the two-ring geometry. Let us consider the following scenario. If we were to stretch a string across the white ring bounded by $R_2$ and $R_3$, in the large $l$ limit, its mass would be of order $\Lambda _{0}=g\sqrt{(M-N)\,l}$, where  $g^{2}=4\pi g_{s}$ \cite{Berenstein:2005aa,Chen:2007gh}. As a matter of fact, $l\,(M-N)$ corresponds to the difference between the weight of a box at the inward corner and the weight of a box at the outward corner along the edge of the Young diagram corresponding to the white ring. The $\Lambda _{0}$ is thus a scale introduced in the $U(M)$ theory. This is analogous to a Coulomb branch vev. When the RG scale $\Lambda \gg \Lambda _{0}$ we would be in the $U(M)$ theory,  whereas for $\Lambda \ll \Lambda _{0}$ it would be effectively described by a $U(N)$ theory. This is analogous to an RG flow on the Coulomb branch. The aforementioned regimes parallel the limits \eqref{limit 1} and \eqref{limit 2}. The coordinate $\left( r^{2}+y^{2}\right) ^{1/2}$ thus seems to geometrize the scale $\Lambda$. Moreover, since the rings are sequenced along the radial direction of the phase space, integrating out the modes near the outer ring would correspond to integrating out the modes that are heavier in the phase space. This qualitative description deserves further investigation.

\subsection{Three-corner backgrounds and compatibility}

\label{sec: compatibility condition}

\begin{figure}[h]
\centering
\includegraphics[trim=0cm 2cm 0cm 2cm, clip=true,scale=0.3]{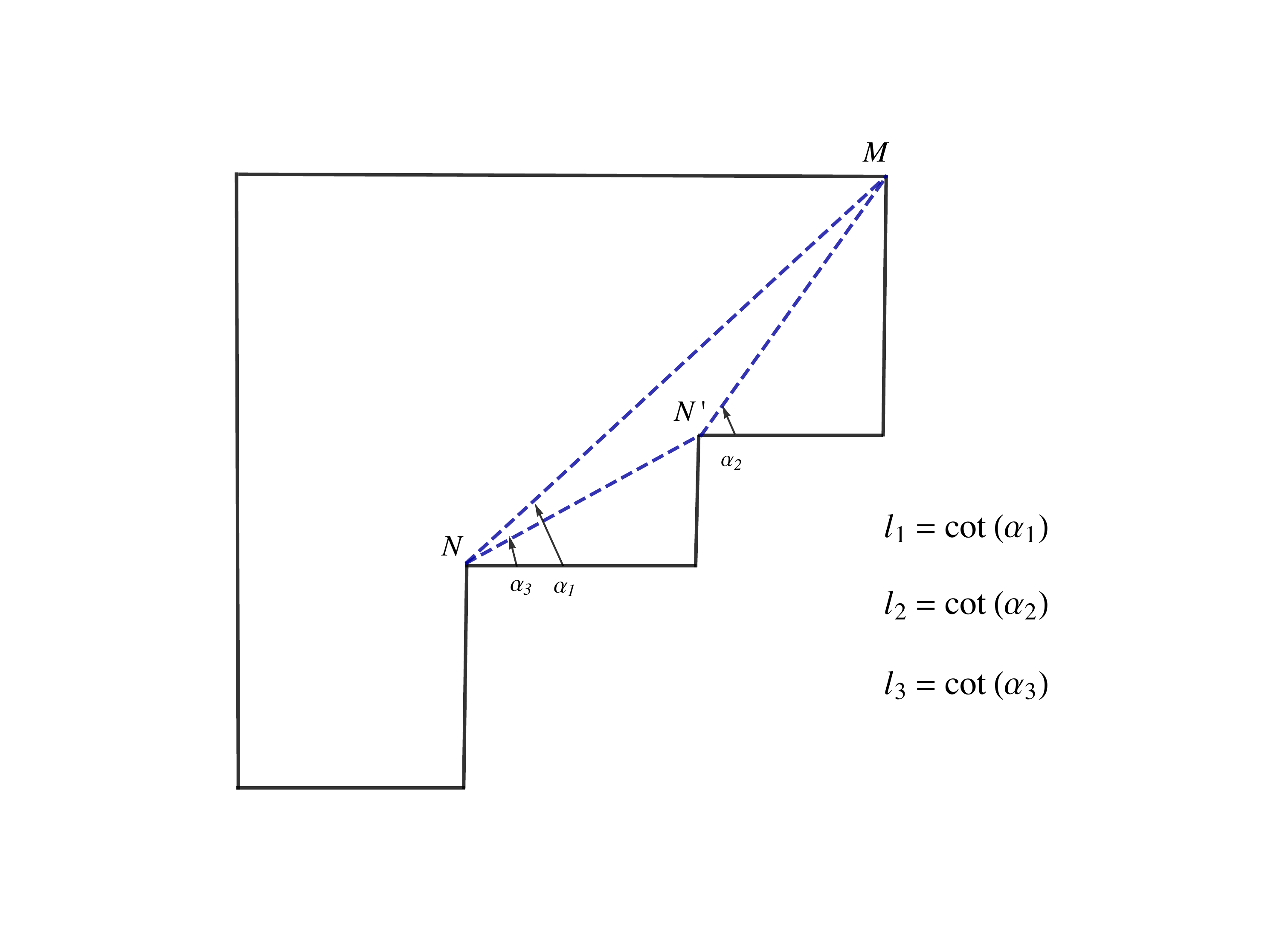}
        \caption{Young diagrams corresponding to three-ring geometries. The angles corresponding to the axial distances must be small for the transition probabilities to match those of $\mathbb{YB}$. }\label{three angles}
\end{figure}

In this subsection, we explain how to understand the compatibility relations \eqref{YBcom} and \eqref{compa gt} in terms of a limit of
transitions on a three-ring background or, equivalently, on a background labeled by a three-corner Young diagram (with three inward corners), as in Fig. \ref{three angles}. Imagine that we are given a two-ring geometry and we start deforming one of the black rings until it splits into two, giving rise to a three-ring geometry. We want to compute transition probabilities analogous to equation \eqref{probabilities} on this kind of backgrounds. Following an analysis similar to the one in subsection \ref{sec: limits}, it is easy to see that the geometry dual to such diagram corresponds to a domain wall with three different AdS regimes, the $(AdS\times S)_{M}$, $(AdS\times S)_{N'}$ and $(AdS\times S)_{N}$. This is a new feature here. As before, we can use the correspondence between  (\ref{GT_embedding_01}) and (\ref{integersequence}) to associate this background with the Gelfand-Tsetlin embeddings of the form
\begin{equation}
U(M)\supset U(N') \supset U(N).
\end{equation}
In this case, the radii of the corresponding three-ring bubbling geometry are given by
\begin{eqnarray}
\tilde{R}_{1} =\sqrt{(l_{M}+1)M}\hspace{3mm}&\;&\hspace{3mm}\tilde{R}_{2} =\sqrt{l_{M}M+N^{\prime }},\nonumber\\
\tilde{R}_{3} =\sqrt{(l_{N^{\prime }}+1)N^{\prime }}\hspace{3mm}&\;&\hspace{3mm}\tilde{R}_{4} =\sqrt{l_{N^{\prime }}N^{\prime }+N},\nonumber\\
\tilde{R}_{5} =\sqrt{(l_{N}+1)N}\hspace{5mm}&\;&\hspace{3mm}\tilde{R}_{6} =\sqrt{l_{N}N}.
\end{eqnarray}

Let us introduce some notation which will be convenient for calculations on backgrounds with more than two corners.
Let $B$ be a Young diagram with $k$ inward corners. We associate to $B$ a map  $ B[\mu_1,\dots,\mu_k ]$ having $k$ entries which are meant to receive a Young diagram each, where the $\mu_1,\dots,\mu_k$ are Young diagrams. In the following, we will denote by $\boldsymbol{\cdot}$ the empty Young diagram. The map works as follows, if we plug a Young diagram $\mu$ into the $i$-th entry the output is the Young diagram $B$ with $\mu$ attached to its $i$-th inward corner, counting downwards. For example, if $k=2$, the objects defined in section \ref{sec: multi-graviton transitions} can be written as
\begin{equation}
B^{\mu}= B[\mu,\boldsymbol{\cdot}]\hspace{5mm}\text{and}\hspace{5mm} B_{\nu}^{\nu'}= B[\nu', \nu].
\end{equation}
 We define also the transition probabilities
\begin{equation}
 P[\mu_1,\mu_2,\dots \,|\, \nu_1,\nu_2,\dots]= {\cal P}\left(B[\mu_1,\mu_2,\dots  ]\rightarrow B[\nu_1,\nu_2,\dots ]\right).
\end{equation}
Hereafter, placing a hat over one of the entries of these probabilities will mean that we are summing over it, for example
\begin{equation}
 P[\mu_1,\mu_2,\dots \,|\,\widehat{\nu_1},\nu_2,\dots]= \sum_{\nu_1}  P[\mu_1,\mu_2,\dots \,|\, \nu_1,\nu_2,\dots]\,.
\end{equation}
With this notation the hook-shaped transition probabilities \eqref{MG} and \eqref{probabilities} read
\begin{equation}
P_{\nu }^{\mu \nu^{\prime }}= P[\mu,\boldsymbol{\cdot}\,|\,\nu',\nu]\hspace{5mm}\text{and}\hspace{5mm}
P_{\nu }^{\mu}= P[\mu,\boldsymbol{\cdot}\,|\, \widehat{\nu'},\nu].
\end{equation}

Now let us consider the following process. Let us start with an excitation $\mu$  at the $M$ corner of the diagram in Fig. \ref{three angles}, which then decays into another excitation $\nu_1$ at the $M$ corner and an excitation $\nu$ at the $N$ corner. Just as in the hook shaped diagrams, we are interested in tracing out the $\nu_1$. Hence, we study the probability
\begin{equation}
P[\mu,\boldsymbol{\cdot},\boldsymbol{\cdot} \,|\,\widehat{\nu_1},\boldsymbol{\cdot},\nu].
\end{equation}
In between these two corners lies corner $N'$, thus we can write the completeness relation
\begin{equation}\label{probs three ring}
 P[\mu,\boldsymbol{\cdot},\boldsymbol{\cdot} \,|\,\widehat{\nu_1},\boldsymbol{\cdot},\nu] = \sum_{\nu'}  P[\mu,\boldsymbol{\cdot},\boldsymbol{\cdot} \,|\,\widehat{\nu_2},\nu', \boldsymbol{\cdot}]\,  P[\boldsymbol{\cdot}, \nu',\boldsymbol{\cdot} \,|\,\boldsymbol{\cdot},\widehat{\nu_3},\nu]\,.
\end{equation}
We saw in Eq. \eqref{final2} in the limit when the corners are well separated the transition probabilities $P[\mu,\boldsymbol{\cdot}\,|\, \widehat{\nu'},\nu]$ correspond to the distributions ${}^{\mathbb{YB}}\Lambda _{r}^{r'}(\mu,\nu )$ of the Young bouquet. Something analogous holds for the transitions appearing in \eqref{probs three ring}, namely
\begin{eqnarray}
  {}^{\mathbb{YB}}\Lambda _{r_{N}}^{r_M}(\mu,\nu ) &=&  \lim_{l_1\to\infty} P[\mu,\boldsymbol{\cdot},\boldsymbol{\cdot} \,|\,\widehat{\nu_1},\boldsymbol{\cdot},\nu], \nonumber\\
 {}^{\mathbb{YB}}\Lambda _{r_{N'}}^{r_M}(\mu,\nu' ) &=&  \lim_{l_2\to\infty} P[\mu,\boldsymbol{\cdot},\boldsymbol{\cdot} \,|\,\widehat{\nu_2},\nu', \boldsymbol{\cdot}],\\
 {}^{\mathbb{YB}}\Lambda _{r_{N}}^{r_{N'}}(\nu',\nu ) &=& \lim_{l_3\to\infty} P[\boldsymbol{\cdot}, \nu',\boldsymbol{\cdot} \,|\,\boldsymbol{\cdot},\widehat{\nu_3},\nu], \nonumber
\end{eqnarray}
where the lengths $l_1$, $l_2$ and $l_3$ are defined in Fig. \ref{three angles}, while $r_M/r_N=M/N$, and so on. The
limits $l_i \gg1$ correspond to taking small angles $\alpha_i$. In fact, $l_2,\, l_3 \gg 1$ guarantee that $l_1 \gg 1$, so it is enough to demand\footnote{In the bubbling plane this condition means that the three rings are well separated.} $\alpha_2,\,\alpha_3\ll 1$. The outcome is that in this regime the completeness relation
\eqref{probs three ring} reproduces the compatibility condition
\begin{equation}
{}^{\mathbb{YB}}\Lambda _{r_{N}}^{r_M}(\mu,\nu )= \sum_{\nu'}\;{}^{\mathbb{YB}}\Lambda _{r_{N'}}^{r_M}(\mu,\nu' ){}^{\mathbb{YB}}\Lambda _{r_{N}}^{r_{N'}}(\nu',\nu ).
\end{equation}
Furthermore, using the $\mathbb{YB}/\mathbb{GT}$ duality \eqref{sim}, we have
\begin{equation}
{}^{\mathbb{GT}}\Lambda _{N^{\prime }}^{M}{}^{\mathbb{GT}}\Lambda
_{N}^{N^{\prime }}={}^{\mathbb{GT}}\Lambda _{N}^{M}, \label{compa gt_02}
\end{equation}%
in the shorthand notation \eqref{compa short}.

\section{Discussion}

\label{sec: discussion}

In this paper we studied an identity relating the branching graphs corresponding to the unitary and symmetric groups in the context of the gauge/gravity correspondence. We proposed to give this identity a physical interpretation. We found that the Young Bouquet distribution in the identity is produced by the square of three point functions, under a natural limit. Moreover, the transition processes in the spacetime that are dual to the processes described by the above three point functions, can be seen as the transitions of multigraviton states in certain domain wall like backgrounds. This graph identity, namely the $\mathbb{YB}$/$\mathbb{GT}$ correspondence, is similar in nature to the Schur-Weyl duality. The latter has been an insightful instrument for the elucidation of the gauge/string correspondence. It has been our guiding physical motivation to employ the $\mathbb{YB}$/$\mathbb{GT}$ correspondence in a similar fashion.

The probabilities that match the information pertaining to the Young-Bouquet graph furnish new observables in the spacetime. These new observables are transition probabilities of the processes in the bulk of the spacetime. We have shown that the background in which these processes take place, in the dual gravitational description, should involve bubbling geometries with multi-ring structures. We claim that these observables live naturally in the bulk and capture quantum gravity effects. It would be good to perform a direct spacetime computation of these observables to confirm our interpretations in the gravity description.

We have computed these observables from correlation functions, at leading order in $N$ and $M$, and it would be interesting to explore subleading corrections. We would like to explore how these corrections reflect upon the structure of branching graphs. A hint as to how to possibly generalize the $\mathbb{YB}$/$\mathbb{GT}$ identity for finite $N$ can be found by looking at the $\mathbb{GT}$ side of the equation, since the conserved charges corresponding to this side were already constructed for finite $N$ \cite{Diaz:2013gja,Diaz:2014ixa}.

Identities similar to $\mathbb{YB}$/$\mathbb{GT}$ for the orthogonal and symplectic groups must exist.
The corresponding finite group will not be ordinary symmetric groups but wreath products $S_n[S_2]$.
They could be investigated from the AdS/CFT correspondence in a similar way as in this work. The groups
can be identified with the gauge groups of $\mathcal{N}=4$ gauge theory, and the $\mathbb{GT}$ should match the eigenvalues
of the embedding charges for the orthogonal and symplectic cases already found in \cite{Diaz:2013gja}. Moreover,
the duality relates the gauge theories with those gauge groups to the type II superstring theory in $AdS_5\times \mathbb{RP}^5$ backgrounds \cite{Witten2,Mukhi:2005cv}. Schur technology for these groups has already been developed \cite{CDD1,CDD2,G1,G2}, and can be used to compute the transition probabilities through the appropriate three-point functions.

Finally, the four-point functions considered in the paper deserve further investigation. As observables they contain more information than the three-point functions, and have a natural interpretation in scattering of multigraviton states in the domain wall like backgrounds that we are considering. Although these observables capture more complicated processes, they also take a compact form depending only on the Littlewood-Richardson coefficients and the dimensions of the irreps. They are associated with restrictions $S_m\supset S_n\times S_{m-n}$ instead of restrictions $S_m\supset S_n$ corresponding to $\mathbb{Y}$. This suggests that the four-point functions might be associated with probabilities coming from some composition of Young graphs. It would be interesting to find their precise relation.

\section*{Acknowledgments}

We would like to thank D. Berenstein, E. O. Colgain, D. Correa, S. Das, R. de Mello Koch, M. Hanada, T. Harmark, A. Jevicki, Y. Kimura, S. Ramgoolam, R. Suzuki, M. Walton and S.-T. Yau for correspondence or discussion. The research of PD is supported by the Natural Sciences and Engineering Research Council of Canada and the University of Lethbridge. The research of HL is supported in part by Center of Mathematical Sciences and Applications, and by NSF grant DMS-1159412, NSF grant PHY-0937443 and NSF grant DMS-0804454. The research of AVO is supported by the University Research Council of the University of the Witwatersrand. AVO thanks the Galileo Galilei Institute for Theoretical Physics for the hospitality and the INFN for partial support during the completion of this work.

\appendix

\section{Four point functions and scattering of multigraviton states}

\label{sec: four point and multigraviton}
\begin{figure}[h!]
\centering
\includegraphics[trim=0cm 3.4cm 0cm 3.2cm, clip=true,scale=0.38,bb=0 90 960 400]{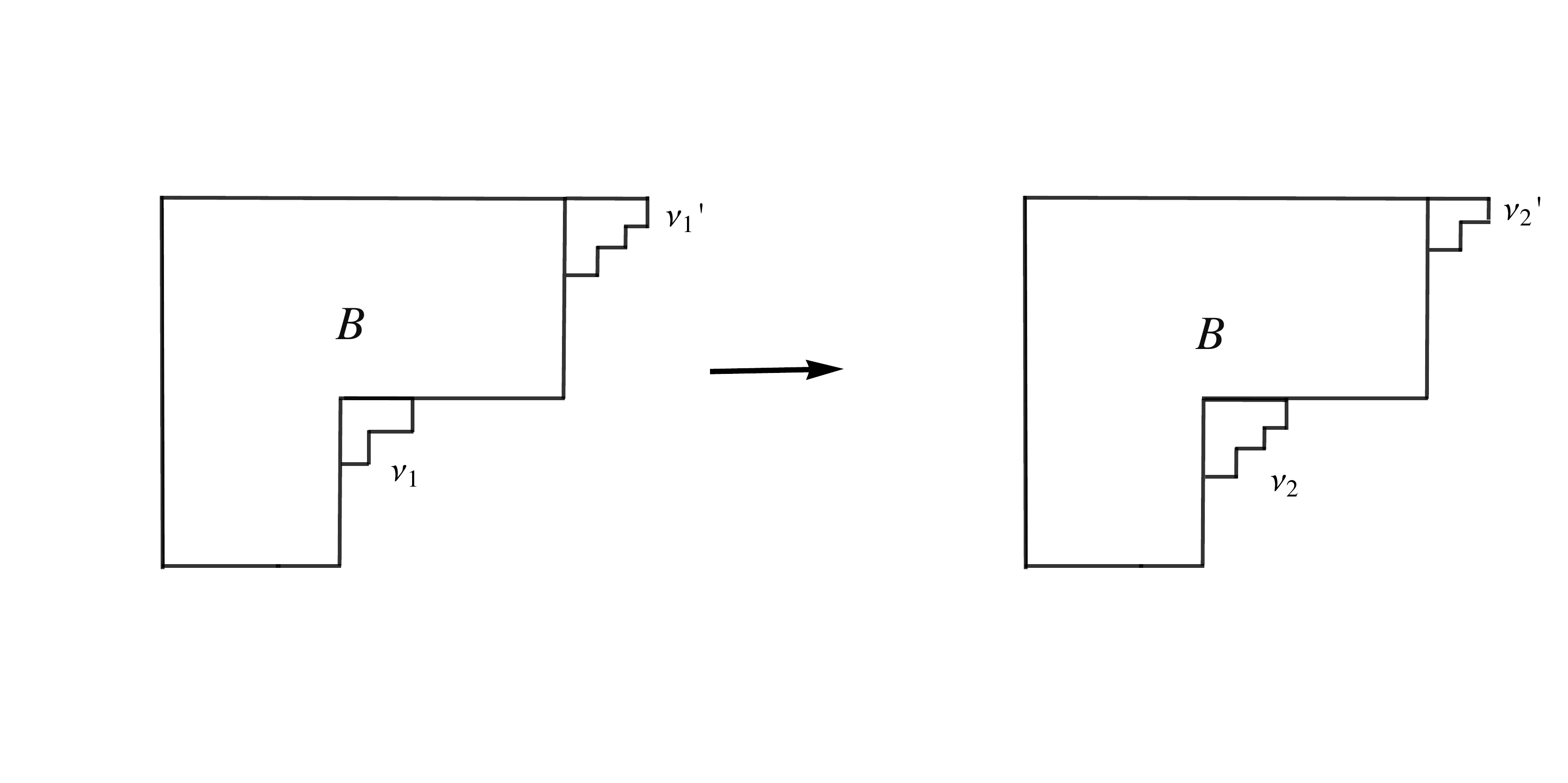}
        \caption{Scattering of a pair of multigraviton states where the conservation of angular momentum leads to the conservation of the number of boxes: $|\nu'_1|+|\nu_1|=|\nu'_2|+|\nu_2|$.}\label{four P}
\end{figure}

We have shown that the probability distributions associated with the Young Bouquet can be obtained from a
precise field theory computation. Our result can be expressed succinctly as
\begin{equation}
{}^{\mathbb{YB}}\Lambda _{r}^{r^{\prime }}(\mu,\nu )=\lim_{l\to\infty}P_{\nu }^{\mu}.
\end{equation}
After showing that, we proceeded to give a gravitational interpretation to these quantities under the light of the gauge-gravity correspondence. The claim is that the Young Bouquet's probability distributions secretly encode the probabilities of certain transition processes and interactions in quantum gravity. Recall that $P_{\nu }^{\mu}$ corresponds to the (square of the) trace of three-point functions displayed in Eq. \eqref{probabilities}. Nothing prevents us from considering other kinds of more involved processes, and one might wonder whether some of these could suggest new insights into the structure of the branching graphs of groups. With this motivation in mind, we consider transition probabilities of four  excitations in this appendix. In the line of subsection \ref{holointer}, we can interpret holographically these correlators as probabilities of a multi-graviton scattering process. This is shown in Fig. \ref{four P}.

Concretely, let us compute the transition probability
\begin{equation}
P_{\nu_1\,\nu_2 }^{\nu'_1\,\nu'_2 }={\cal P}(B_{\nu'_1 }^{\nu_1}\rightarrow B_{\nu'_2 }^{\nu_2}),\label{P4}
\end{equation}
displayed in Fig. \ref{four P}. In the dual picture, that would correspond to computing the scattering
 between multigraviton excitations ($\nu _{1}^{\prime },$ $\nu _{1}$) with angular momenta ($n_{1}^{\prime
},n_{1}$) and multigraviton excitations ($\nu _{2}^{\prime },$ $\nu _{2}$)
with angular momenta ($n_{2}^{\prime },n_{2}$) in the two-ring background. Clearly, we must demand conservation of angular momentum, hence the number of boxes in the Young diagrams must satisfy
\begin{equation}\label{cons}
|\nu _{1}|+|\nu _{1}^{\prime }|=|\nu _{2}|+|\nu _{2}^{\prime }|.
\end{equation}
In practice, we calculate
\begin{equation}
P_{\nu_1\,\nu_2 }^{\nu'_1\,\nu'_2 }\;=\sum_{\mu \vdash |\nu _{1}|+|\nu _{1}^{\prime }|}|\langle
{} \chi _{\mu }^{\dagger }(Y)\chi _{B_{\nu _{1}}^{\nu _{1}^{\prime
}},(B,\mu )}(Z,Y)\chi _{\mu }(Y)\chi _{B_{\nu _{2}}^{\nu _{2}^{\prime
}},(B,\mu )}^{\dagger }(Z,Y) \rangle|_{norm}^{2} \, ,\label{P4_02}
\end{equation}
where the four-point functions\footnote{By conformal transformation, we can extract the four point functions by conformally transform them from $R^4$ to $R \times S^3$. It is easiest to arrange the four points, after transformation, to be along the $R$ direction. For the correlator in Eq. (\ref{P4_02}), the positions of the four operators are $p'_1, p_2, p_1, p'_2$ respectively along the $R$ direction. Since the fields are in definite representations of the conformal group $SO(4,2)$, the spatial dependence of the correlator are transformed back from $R \times S^3$ to $R^4$ by keeping tracking of which irreps of $SO(4,2)$ that the fields are in.} must be appropriately normalized. In large $N$ and $M$, the operator product expansion implies that
the above correlators factorize as
\begin{eqnarray}
&&\left\langle {}\right. \chi _{\mu }^{\dagger }(Y)\chi _{B_{\nu _{1}}^{\nu
_{1}^{\prime }},(B,\mu )}(Z,Y)\chi _{\mu }(Y)\chi _{B_{\nu _{2}}^{\nu
_{2}^{\prime }},(B,\mu )}^{\dagger }(Z,Y)\left. {}\right\rangle _{norm}
 \\
&=&\left\langle {}\right. \chi _{\mu }^{\dagger }(Y)\chi _{B}^{\dagger
}(Z)\chi _{B_{\nu _{1}}^{\nu _{1}^{\prime }},(B,\mu )}(Z,Y)\left.
{}\right\rangle _{norm.}\left\langle {}\right. \chi _{\mu }(Y)\chi
_{B}(Z)\chi _{B_{\nu _{2}}^{\nu _{2}^{\prime }},(B,\mu )}^{\dagger}(Z,Y)\left. {}\right\rangle _{norm}.\notag
\end{eqnarray}
Therefore, we find
\begin{equation}
P_{\nu_1\,\nu_2 }^{\nu'_1\,\nu'_2 }\;=\sum_{\mu \vdash |\nu _{1}|+|\nu _{1}^{\prime }|}P_{\nu _{1}}^{\mu \nu
_{1}^{\prime }}P_{\nu _{2}}^{\mu \nu _{2}^{\prime }}.
\end{equation}
Now, introducing the variable
\begin{equation}
x=\frac{r}{r^{\prime }}\left( \frac{l}{l+1}\right),
\end{equation}
and using \eqref{l}, we rewrite Eq. \eqref{Pmnn} as
\begin{equation}
P_{\nu }^{\mu \nu ^{\prime }}=x^{n}(1-x)^{m-n}\frac{m!}{n!(m-n)!}g(\mu ;\nu ,\nu ^{\prime })\frac{\dim
_{\nu ^{\prime }}\dim _{\nu }}{\dim _{\mu }}.
\end{equation}
Hence, we obtain
\begin{eqnarray}
P_{\nu_1\,\nu_2 }^{\nu'_1\,\nu'_2 } &=&x^{n_{1}+n_{2}}(1-x)^{n_{1}^{\prime }+n_{2}^{\prime }}\binom{%
n_{1}+n_{1}^{\prime }}{n_{1}}\binom{n_{2}+n_{2}^{\prime }}{n_{2}}\dim _{\nu
_{1}}\dim _{\nu _{1}^{\prime }}\dim _{\nu _{2}}\dim _{\nu _{2}^{\prime }}
\notag \\
&&\sum_{\mu \vdash |\nu _{1}|+|\nu _{1}^{\prime }|}\frac{g(\mu ;\nu _{1},\nu
_{1}^{\prime })g(\mu ;\nu _{2},\nu _{2}^{\prime })}{\left( \dim _{\mu
}\right) ^{2}}\;.  \label{P_03}
\end{eqnarray}
This formula has two kinds of contributions, one involving just $\{n_{1}^{\prime },n_{1}$,$n_{2}^{\prime },n_{2}\}$, which corresponds to a kinematic
factor, and another one containing only $\{\nu _{1}^{\prime },\nu _{1},\nu_{2}^{\prime },\nu _{2}\}$, that distinguishes excitations with the same number of boxes but labeled by different Young diagrams. The different Young diagrams correspond to different combinations in the multi-graviton states.

As said before these processes can be interpreted as scattering of multi-gravitons in the dual picture. In this picture, the center of mass energy of the multi-graviton scattering process is given by $m=|\mu|$. Conservation of angular momentum \eqref{cons} implies that $m=n_1+n_1'=n_2+n_2'$, where $m$ is the total number of boxes of the small Young diagrams. Moreover, in the large $l$ limit
\begin{equation}
x=\frac{N}{M}\left(1-{\cal O}(\frac{1}{l})\right).
\end{equation}
Therefore, in the large $M$ limit the scattering probability \eqref{P_03} behaves as
\begin{equation}
P_{\nu_1\,\nu_2 }^{\nu'_1\,\nu'_2 }\sim\,\frac{1}{M^{2m}}\; .
\end{equation}
The above equation keeps track of the dependence on the number of powers of $1/M$ from Eq. \eqref{P_03}, in $1/M$ expansions.
This suggests that the scattering process depicted in Fig. \ref{four P}
captures quantum effects due to gravitational interactions. For related
discussion on the transition amplitudes in the gravity, and on four point
correlation functions with the large dimension operators, see for example
\cite{Brown:2006zk,Lin:2012ey}.

\end{document}